%% file: revshv.tex
\documentclass[a4paper]{iopart}
\usepackage{a4wide}
\usepackage{amssymb,amsfonts,latexsym,stmaryrd}
\usepackage[PostScript=dvips]{diagrams}
\usepackage{QED}
\usepackage{euscript}
\usepackage{tikz,ifdraft}
\usetikzlibrary{arrows,positioning,matrix,backgrounds,calc,fit} 
\usepackage{url}

\newtheorem{theorem}{Theorem}[section]

\newtheorem{proposition}[theorem]{Proposition}

\newcommand{\beqa}{\begin{eqnarray*}}
\newcommand{\eeqa}{\end{eqnarray*}\par\noindent}

\newcommand{\BB}{\mathbf{B}}

\newcommand{\true}{\mathsf{true}}
\newcommand{\false}{\mathsf{false}}

\newcommand{\XX}{\mathcal{X}}
\newcommand{\SSS}{\mathcal{S}}

\newcommand{\lrarr}{\longrightarrow}
\newcommand{\rarr}{\rightarrow}
\newcommand{\HH}{\mathcal{H}}

\newcommand{\ket}[1]{{|} #1\rangle}
\newcommand{\bra}[1]{\langle #1{|}}
\newcommand{\ie}{\textit{i.e.}~}
\newcommand{\CC}{\mathcal{C}}

\newcommand{\DD}{\mathcal{D}}
\newcommand{\EE}{\mathcal{E}}

\newcommand{\Set}{\mathbf{Set}}

\newcommand{\Real}{\mathbb{R}}

\newcommand{\id}{\mathsf{id}}

\newcommand{\Pow}{\mathcal{P}}

\newcommand{\AND}{\; \wedge \;}
\newcommand{\OR}{\; \vee \;}

\newcommand{\da}{{\downarrow}}
\newcommand{\vn}{\varnothing}

\newcommand{\card}[1]{|#1|}

\newarrow{Eq}=====
\newarrow{Dash}.....
\newarrow{mon}>--->
\newarrow{inc}C--->

\diagramstyle[newarrowhead=vee,newarrowtail=vee]

\newcommand{\op}{\mathsf{op}}

\newcommand{\fres}{|}

\newcommand{\UU}{\mathcal{U}}

\newcommand{\Rpos}{\Real_{\geq 0}}

\newcommand{\Bool}{\mathbb{B}}

\newcommand{\supp}{\mathsf{supp}}
\newcommand{\DR}{\mathcal{D}_{R}}

\newcommand{\DB}{\mathcal{D}_{\Bool}}
\newcommand{\DP}{\mathcal{D}_{\Rpos}}
\newcommand{\DReal}{\mathcal{D}_{\Real}}

\newcommand{\hcl}{h^{\lambda}_{C}}
\newcommand{\hcpl}{h^{\lambda}_{C'}}

\newcommand{\hl}{h_{\Lambda}}

\newcommand{\SP}{\SSS}

\newcommand{\RR}{\mathcal{R}}

\newcommand{\XOR}{\mathsf{ONE}}
\newcommand{\Proj}{\mathbf{P}}
\newcommand{\ee}{\mathbf{e}}
\newcommand{\Prei}{\Proj_{\ee_i}}

\newcommand{\Prej}{\Proj_{\ee_j}}
\newcommand{\Pnj}{\Proj_{\ee_j}^{\bot}}
\newcommand{\Aei}{A_{\ee_{i}}}
\newcommand{\Aej}{A_{\ee_{j}}}
\newcommand{\Aek}{A_{\ee_{k}}}
\newcommand{\Aeone}{A_{\ee_{1}}}

\newcommand{\amp}{\alpha_{m}^{0}}
\newcommand{\amn}{\alpha_{m}^{1}}
\newcommand{\Prmp}{\Proj_{m}^{0}}
\newcommand{\Prmn}{\Proj_{m}^{1}}

\newcommand{\Ps}{\Proj_{s}}
\newcommand{\Pms}[1]{\Proj_{m_{#1}}^{s(m_{#1})}}

\newcommand{\rmap}[2]{\mathsf{res}^{#1}_{#2}}
\newcommand{\MM}{\mathcal{M}}
\newcommand{\MG}{\MM_{G}}
\newcommand{\MB}{\mathbf{M}}
\newcommand{\xx}{\mathbf{X}}
\newcommand{\vv}{\mathbf{V}}
\renewcommand{\emph}{\textbf}
\newcommand{\mxi}{X^{(i)}}
\newcommand{\mxj}{X^{(j)}}
\newcommand{\mxk}{X^{(k)}}
\newcommand{\myj}{Y^{(j)}}
\newcommand{\myk}{Y^{(k)}}
\newcommand{\myi}{Y^{(i)}}
\newcommand{\hlm}{h^{\lambda}_{m}}
\newcommand{\hlim}{h^{\lambda}_{m_i}}
\newcommand{\hlx}{h^{\lambda}_{X}}
\newcommand{\ua}{{\uparrow}}

\newcommand{\UUj}{\UU^{(j)}}
\newcommand{\EEj}[1]{\EE^{(#1)}}
\newcommand{\ej}[1]{e^{(#1)}}

\newcommand{\tus}{v_{U,s}}
\newcommand{\tusp}{v_{U',s'}}
\newcommand{\tusq}{v_{U'',s''}}
\newcommand{\vus}{\mathbf{v}_{U,s}}

\newcommand{\mus}{\mu_{U,s}}
\newcommand{\musp}{\mu_{U',s'}}
\newcommand{\musq}{\mu_{U'',s''}}
\newcommand{\Rt}{\Real^t}
\newcommand{\sm}[1]{s_{m := #1}}

\begin{document}

\title{The Sheaf-Theoretic Structure Of Non-Locality and Contextuality}
\author{Samson Abramsky$^1$ and Adam Brandenburger$^2$}
\address{$^1$Department of Computer Science, University of Oxford, Wolfson Building, Parks Road, Oxford OX1 3QD, UK}
\address{$^2$Stern School of Business, New York University, 44 West Fourth Street, New York, NY 10012}
\eads{\mailto{samson@comlab.ox.ac.uk}, \mailto{adam.brandenburger@stern.nyu.edu}}

\begin{abstract}
We use the mathematical language of sheaf theory to give a unified treatment of non-locality and contextuality, in a setting which generalizes the familiar probability tables used in non-locality theory to arbitrary \textit{measurement covers}; this includes Kochen-Specker configurations and more. We show that contextuality, and non-locality as a special case, correspond exactly to \textit{obstructions to the existence of global sections}.
We describe a linear algebraic approach to computing these obstructions, which allows a systematic treatment of arguments for non-locality and contextuality. 
We distinguish a proper hierarchy of strengths of no-go theorems, and show that three leading examples --- due to Bell, Hardy, and Greenberger, Horne and Zeilinger, respectively --- occupy successively higher levels of this hierarchy.  
A general correspondence is shown between the existence of local hidden-variable realizations using negative probabilities, and no-signalling; this is based on a result showing that the linear subspaces generated by the non-contextual and no-signalling models, over an arbitrary measurement cover, coincide. Maximal non-locality is generalized to maximal contextuality, and characterized in purely qualitative terms, as the non-existence of global sections in the support. 
A general setting is developed for Kochen-Specker type results, as generic, model-independent proofs of maximal contextuality, and a new combinatorial condition is given, which generalizes the `parity proofs' commonly found in the literature. We also show how our abstract setting can be represented in quantum mechanics. This leads to a strengthening of the usual no-signalling theorem, which shows that quantum mechanics obeys no-signalling for arbitrary families of commuting observables, not just those represented on different factors of a tensor product.  
\end{abstract}

\pacs{03.65.Ud,03.65.Fd,03.65.Ca}

\submitto{\NJP}

\maketitle

\section{Introduction}

Non-locality and contextuality are fundamental features of physical theories, which contradict the intuitions underlying classical physics. They are, in particular, prominent features of quantum mechanics, and the goal of the  classic no-go theorems by Bell \cite{bell1964einstein}, Kochen-Specker \cite{kochen1975problem}, et al.~is to show that they are \textit{necessary features} of any theory whose experimental predictions agree with those of quantum mechanics.

Bell's insights into non-locality have been seminal to the current developments in quantum information, where entanglement is viewed as a key informatic resource; and there has also been considerable recent work on experimental tests for contextuality \cite{bartosik2009experimental,kirchmair2009state}.

In the present paper, we study these notions from a novel perspective, which yields  new insights and results.
Our approach has the following notable features:
\begin{itemize}
\item The importance of Bell's theorem and related results is that they apply, not just to quantum mechanics, but to \textit{all} theories with certain structural properties. We introduce a general mathematical setting, completely independent of Hilbert space, which strengthens this feature, and allows results to be proved in considerable generality.

\item We study non-locality and contextuality in a unified setting.
The idea that non-locality can be seen as a particular form of contextuality, and specific results obtaining Bell-type non-locality  from Kochen-Specker configurations, can be found in references such as \cite{heywood1983nonlocality,stairs1983quantum,mermin1990simple}.
An important recent contribution in this direction is \cite{cabello2010non}, which studies non-contextual inequalities as a generalization of Bell inequalities. 

Our approach focusses on structural aspects. It offers a general, systematic and mathematically robust setting in which non-locality and contextuality are treated in a unified fashion; our definitions and results specialize to yield standard formulations of either as special cases, but subsume both.

\item We use the mathematics of \emph{sheaf theory} to analyze the structure of non-locality and contextuality. Sheaf theory is pervasive in modern mathematics, allowing the passage from local to global \cite{mac1992sheaves}. Starting from a simple experimental scenario, and the kind of probabilistic models familiar from discussions of Bell's theorem, Popescu-Rohrlich boxes \cite{popescu1994quantum}, etc., we show that there is a very direct, compelling formalization of these notions in sheaf-theoretic terms.
Moreover, on the basis of this formulation, we show that the phenomena of non-locality and contextuality can be characterized precisely  in terms of \emph{obstructions to the existence of global sections}. We give linear algebraic methods for computing these obstructions.
\end{itemize}

These ideas lead in turn to a number of novel insights into non-locality and contextuality:
\begin{itemize}
\item We are able to distinguish three \emph{strengths} of degree of non-locality: standard \emph{probabilistic non-locality}, exhibited by the original example of Bell; \emph{possibilistic non-locality}, exemplified by the well-known \emph{Hardy model} \cite{hardy1993nonlocality}; and \emph{strong contextuality}. These three properties form a strict hierarchy; strong contextuality implies possibilistic  non-locality, which implies probabilistic non-locality, but the converse implications fail.
In fact, we show that the Bell model is probabilistically but not possibilistically non-local; the Hardy model is possibilistically non-local but not strongly contextual; and the \emph{GHZ models} \cite{greenberger1989going}, for all numbers of parties greater than $2$, are strongly contextual. Thus we have a hierarchy
\[ \mbox{Bell} < \mbox{Hardy} < \mbox{GHZ} . \]
Moreover, Ray Lal has shown (private communication) that the only bipartite no-signalling devices satisfying strong contextuality are the PR boxes, thus giving a new characterization of these super-quantum devices.

\item We show that strong contextuality is equivalent to a quantitative notion of \emph{maximal contextuality}, which has been studied in the special case of Bell-type scenarios as \emph{maximal non-locality}. We use this equivalence to characterize maximal contextuality, and in particular maximal non-locality, in terms of a \emph{boolean satisfiability problem} naturally associated with a probabilistic model, for the case of dichotomic measurements, and more generally in terms of a \emph{constraint satisfaction problem}.

\item We  apply our linear algebraic methods for constructing global sections to the issue of giving local hidden-variable realizations using \emph{negative probabilities} \cite{wigner1932quantum,Dirac42,moyal1949quantum,feynman1987negative}.
We show that there is an equivalence between the existence of such realizations, and the \emph{no-signalling} property.

\item We give a general perspective on Kochen-Specker type theorems as generic (model-independent) proofs of strong contextuality. We show the general combinatorial structure of these results, and make connections to graph theory, leading to a notion of \emph{Kochen-Specker graphs}, defined in purely graph-theoretic terms.

\item   We prove a general result (Theorem~\ref{equivth}) which shows a strict equivalence between the realization of a system by a \emph{factorizable hidden-variable model}, and the existence of a global section which glues together a certain compatible family on a presheaf. 
Factorizability is a general property, which subsumes both Bell-locality and a form of non-contextuality at the level of distributions as special cases.
This means that the whole issue of  non-locality and contextuality can be translated into a canonical mathematical form, in terms of obstructions to the existence of certain global sections. This opens up the possibility of applying the powerful methods of sheaf theory to studying the structure of these notions.

\item We show in detail how the abstract setting we use can be represented in quantum mechanics; hence our results apply to all the standard situations. One interesting point which emerges from this is that the property of \emph{compatibility} of a family of sections on a presheaf corresponds to a form of \emph{no-signalling} \cite{ghirardi1980general}. This form of no-signalling subsumes, but is more general than, the usual notion; it applies to arbitrary families of commuting observables, not just those represented on different factors of a tensor product. We therefore prove a \emph{generalized no-signalling theorem}, showing that quantum mechanics does satisfy this more general property.

\end{itemize}

The remainder of this paper is organized as follows. The basic setting is motivated and laid out in Section~2. The correspondence between global sections and (deterministic) local hidden variables is explained in Section~3. The linear algebraic method for constructing global sections (or determining their non-existence) is presented in Section~4, together with the results relating to the Bell and Hardy models. 
The equivalence between no-signalling and the existence of local hidden-variable realizations with negative probabilities is proved in Section~5. Strong contextuality, the results relating to the GHZ models and the hierarchy between Bell, Hardy and GHZ, and the connections with maximal non-locality, are presented in Section~6.
The general combinatorial structure of Kochen-Specker-type theorems is studied in Section~7.
In Section~8, we prove our general result relating factorizable hidden-variable models to the existence of global sections. Representations in quantum mechanics, and the generalized form of no-signalling, are treated in Section~9.  Section~10 contains a postlude, summarizing what has been done, discussing related work, and describing some further  directions.

The mathematical background needed to read this paper is quite modest.
In particular, only the bare definitions of category and functor are required. A brief appendix reviews these definitions.

\section{The Setting}

\subsection{A Basic Scenario}

Our starting point is the idealized situation depicted in the following diagram.

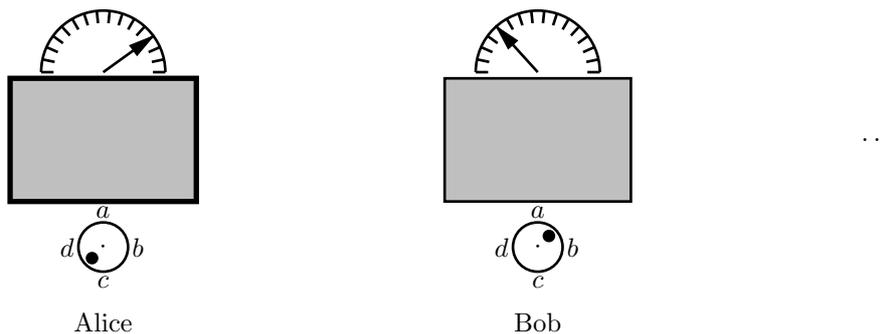
\begin{center}
\input{scenario.tex}
\end{center}

There are several agents or experimenters, who can each select one of several different measurements to perform, and  observe one of several different outcomes. These agents may or may not be spatially separated. When a system is prepared in a certain fashion and measurements are selected, some corresponding outcomes will be observed. These individual occurrences or `runs' of the system are the basic events.
Repeated runs allow relative frequencies to be tabulated, which can be summarized by a probability distribution on events for each selection of measurements. We shall call such a family of probability distributions, one for each choice of measurements,  an \emph{empirical model}.

As an example of such a model, consider the following table.
\begin{center}
\begin{tabular}{ll|ccccc}
A & B & $(0, 0)$ & $(1, 0)$ & $(0, 1)$ & $(1, 1)$  &  \\ \hline
$a$ & $b$ & $1/2$ & $0$ & $0$ & $1/2$ & \\
$a'$ & $b$ & $3/8$ & $1/8$ & $1/8$ & $3/8$ & \\
$a$ & $b'$ & $3/8$ & $1/8$ & $1/8$ & $3/8$ &  \\
$a'$ & $b'$ & $1/8$ & $3/8$ & $3/8$ & $1/8$ & 
\end{tabular}
\end{center}
The intended scenario here is that Alice can choose between measurements $a$ and $a'$, and Bob can choose $b$ or $b'$. Thus the \emph{measurement contexts}  are
\[ \{ a, b \}, \quad \{ a', b\}, \quad \{ a, b' \}, \quad \{ a', b' \} ,\]
and these index the rows of the table.
Each measurement has possible outcomes $0$ or $1$.
The matrix cell at row $(a', b)$ and column $(0,1)$ corresponds to the event where Alice performs $a'$ and observes the outcome $0$, and Bob performs $b$ and observes the outcome $1$.
This can be described by the function
\[ \{ a' \mapsto 0,  \; b \mapsto 1 \} . \]
The cells of the row indexed by $\{ a', b \}$ correspond to the set of functions $O^C$, where $C$ is the measurement context $\{ a', b \}$, and $O = \{ 0, 1 \}$ is the set of outcomes.\footnote{$O^C$ denotes the set of functions from $C$ to $O$. This and a few other set-theoretic notations are explained in the Appendix.}

Each row of the table specifies a probability distribution on events for a given choice of measurements, \ie on the set $O^C$ where the row is indexed by $C$. For example, the event 
\[ \{ a' \mapsto 0, \; b \mapsto 1 \}  \]
is specified to have the probability $1/8$.

The basic ingredients of our formalism will be the \emph{measurement contexts}, the \emph{events}, and the \emph{distributions} on events. A model of a particular measurement scenario will be given by specifying a set of measurements $X$, a family $\MM$ of measurement contexts, and for each context $C \in \MM$, a distribution on the events $O^C$.

We shall now proceed to formalize these ideas. Simple as this setting may seem, it does have significant mathematical structure, which our formalization will enable us to  articulate.

\subsection{Events}

We shall fix a set $X$ of measurements.
We shall also fix  a set $O$ of possible \emph{outcomes} for each measurement.\footnote{We could allow a different set of outcomes for each individual measurement, but we will not need this extra generality.}
Throughout this paper, we shall assume that $X$ and $O$ are finite.
%; we shall comment briefly on generalizations in the final section.

For each set of measurements  $U \subseteq X$, a \emph{section over $U$}
is a function $s : U \rarr O$.
Such a section describes the event in which the measurements in $U$ were performed, and the outcome $s(m)$ was observed for each $m \in U$ .

We shall write $\EE : U \mapsto O^U$ for the assignment of the set of sections over $U$ to each set of measurements $U$.
There is also a natural action by restriction. If $U \subseteq U'$, there is a map
\[ \rmap{U'}{U} : \EE(U') \rarr \EE(U) :: s \mapsto s | U . \]
Note that $\rmap{U}{U} = \id_{U}$, and if $U \subseteq U' \subseteq U''$, then
\[ \rmap{U'}{U} \circ \rmap{U''}{U'} = \rmap{U''}{U} . \]
Altogether, this says that $\EE$ is a \emph{presheaf}, \ie a functor $\EE : \Pow(X)^{\op} \lrarr \Set$.

$\EE$ has an important additional property. Suppose we are given a family of sets $\{ U_i \}_{i \in I}$ with $\bigcup_{i \in I} U_i = U$; \ie the family $\{ U_i \}$ is a \emph{cover} of $U$. Suppose moreover that we are given a family of sections $\{ s_i \in \EE(U_i) \}_{i \in I}$, which is \emph{compatible} in the following sense: for all $i, j \in I$,
\[ s_i | U_i \cap U_j = s_j | U_i \cap U_j . \]
Then there is a unique section $s \in \EE(U)$ such that $s | U_i = s_i$ for all $i \in I$.
This says that we can glue together local data which is compatible in the sense of agreeing on overlaps; moreover, this glued section is uniquely determined.

This gluing property is known as the \emph{sheaf condition}; it  says that $\EE$ is a sheaf, which we shall refer to as the \emph{sheaf of events}.

The fact that this sheaf condition  holds for $\EE$ is quite trivial, since we are simply looking at functions on a discrete space; we can always glue together partial functions which agree on their overlaps, just by taking the union of their graphs.

\subsection{Distributions}

To capture the idea that empirically we observe statistical rather than deterministic behaviour in microphysical systems, we shall consider distributions on events.
It will be advantageous to allow some generality in the notion of distribution we shall consider, by taking the algebra of probabilistic `weights' as a parameter.

A \emph{commutative semiring} is a structure $(R, +, 0, \cdot, 1)$, where $(R, +, 0)$ and $(R, \cdot, 1)$ are commutative monoids, and moreover multiplication distributes over addition:
\[ x \cdot (y + z) = x \cdot y + x \cdot z . \]
There are three main examples of semirings which will be of interest:
the reals
\[ (\Real, +, 0, {\times}, 1), \]
 the non-negative reals
\[ (\Rpos, +, 0, {\times}, 1), \]
and the booleans
\[ \Bool = (\{ 0, 1 \}, \vee, 0, \wedge, 1) . \]
We fix a semiring $R$. Given a set $X$, the \emph{support} of a function $\phi : X \rarr R$ is the set of $x \in X$ such that $\phi(x) \neq 0$. We write $\supp(\phi)$ for the support of $\phi$. An \emph{$R$-distribution} on $X$ is a function  $d : X \rarr R$ which has finite support, and such that
\[ \sum_{x \in X} d(x) = 1 . \]
Note that the finite support condition ensures that this sum is well-defined.
We write $\DR(X)$ for the set of $R$-distributions on $X$.

In the case of the semiring $\Rpos$, this is the set of probability distributions with finite support on $X$.  In the case of the booleans $\Bool$, it is the set of non-empty finite subsets of $X$; thus possibilistic or relational models \cite{abramsky2010relational,fritz2009possibilistic} are also covered.
In the case of the reals $\Real$, it is the set of \emph{signed measures} with finite support, allowing for `negative probabilities' \cite{wigner1932quantum,Dirac42,moyal1949quantum,feynman1987negative}.

Given a function $f : X \rarr Y$, we define
\[ \DR(f) : \DR(X) \rarr \DR(Y) :: d \mapsto [ y \mapsto \sum_{f(x) = y} d(x) ] . \]
This is easily seen to be functorial:
\[ \DR(g \circ f) = \DR(g) \circ \DR(f), \qquad \DR(\id_{X}) = \id_{\DR(X)} \]
so we have a functor $\DR : \Set \lrarr \Set$.\footnote{This functor forms part of the well-known \emph{distribution monad}; see e.g. \cite{jacobs2010convexity} for references.}

We can compose this functor with the event sheaf $\EE : \Pow(X)^{\op} \lrarr \Set$, to form a presheaf $\DR \EE : \Pow(X)^{\op} \lrarr \Set$, which assigns to each set of measurements $U$ the set $\DR (\EE(U))$ of distributions on $U$-sections. It is worth writing out the functorial action of this presheaf explicitly. Given $U \subseteq U'$ we have a map
\[ \DR\EE(U') \rarr \DR\EE(U) :: d \mapsto d | U , \]
where for each $s \in \EE(U)$:
\[ d | U (s) \; := \; \sum_{s' \in \EE(U'), s' | U = s} d(s') . \]
Thus $d | U$ is the \emph{marginal} of the distribution $d$, which assigns to each section $s$ in the smaller context $U$ the sum of the weights of all sections $s'$ in the larger context which restrict to $s$.

\subsection{Measurement Covers}

A crucial point is that it may \textit{not} be possible, in general, to perform all  measurements together.
This is implicit in the idea that each agent makes a choice of measurement from several alternatives; only the measurements which are chosen are actually performed. In the situation where the agents are spatially separated, and the measurements which each performs are localized to their own site, the measurements at the different parts can be performed jointly. In general, we must allow for more complex situations, where compatible sets of measurements may overlap in complicated ways.

We shall now introduce the notion of  measurement cover, which formalizes the idea that only certain measurements can be performed jointly.

A \emph{measurement cover} $\MM$ on the set $X$ of measurements is a family of  subsets of $X$ such that:
\begin{itemize}
\item $\bigcup \MM = X$.
\item $\MM$ is  an  \emph{anti-chain}, \ie   $C, C' \in \MM$ and $C \subseteq C'$ implies $C = C'$.
\end{itemize}

We think of $X$ as a set of labels for the basic measurements in an experiment. A set $C \in \MM$ is a \emph{measurement context}; a set of measurements which can be performed jointly.
We shall focus on the \emph{maximal} compatible sets of measurements, hence the anti-chain condition. Any compatible family of measurements in $X$ will be included in some element of $\MM$.

It should be noted that measurement covers provide a very general way of representing compatibility relationships. Of course, a physical interpretation in particular circumstances will give rise to specific structures of this kind. We shall discuss quantum representations of the formalism in detail in Section~\ref{qrep}. We also discuss the conceptual consequences of our (and related) results concerning compatibility in the Postlude in Section~\ref{postludesec}.

\subsubsection{Bell-Type Scenarios}

We shall now describe a particular class of measurement covers which arises in the formulation of Bell-type theorems on non-locality, and in the study of PR-boxes and other non-local devices.

Consider a  disjoint family $\{ X_i \}_{i \in I}$. We think of $I$ as labelling the parts of a system, which may be space-like separated; $X_i$ is the set of basic measurements which can be performed at part $i$. We form the disjoint union $X$ of this family. We define $\MM$ to be those subsets of $X$  containing exactly one measurement from each part. Thus we regard measurements performed in different parts of the system as compatible, but do not allow for compatible measurements in the same part. 

\subsubsection{Kochen-Specker-Type Scenarios}
\label{KSscensubsec}

Measurement covers are general enough to cover the situations arising in Kochen-Specker style proofs of contextuality, as well as the Bell-type scenarios for non-locality.

Consider the set $X = \{ m_1, \ldots , m_{18} \}$, and the measurement  cover $\MM$ whose elements are the columns of the following table:
\begin{center}
\begin{tabular}{|c|c|c|c|c|c|c|c|c|} \hline
$m_1$ & $m_1$ & $m_8$ & $m_8$ & $m_2$ & $m_9$ & $m_{16}$ & $m_{16}$  & $m_{17}$ \\ \hline
$m_2$ & $m_5$ & $m_9$ & $m_{11}$ & $m_5$ & $m_{11}$ & $m_{17}$ & $m_{18}$ & $m_{18}$  \\ \hline
$m_3$ & $m_6$ & $m_3$ & $m_7$ & $m_{13}$ &  $m_{14}$ & $m_4$ & $m_6$ & $m_{13}$  \\ \hline
$m_4$ & $m_7$ & $m_{10}$ & $m_{12}$ & $m_{14}$  & $m_{15}$ & $m_{10}$ & $m_{12}$ & $m_{15}$  \\ \hline
\end{tabular}
\end{center}
The importance of this example is that it can be realized by unit vectors in $\Real^{4}$, such that each measurement context $C$ in $\MM$ is an orthogonal set of vectors. This structure is used in the 18-vector proof of the Kochen-Specker theorem  in \cite{cabello1996bell}.

We shall discuss proofs of the Kochen-Specker theorem in detail later in the paper; from the point of view of the abstract, `logical' structure in Section~\ref{abskssec}, and as regards the interpretation in quantum mechanics in Section~\ref{KSqrep}.

\subsection{Empirical Models}

We shall now show how the intuitive scenario described at the beginning of this section can be captured formally, using the mathematical structure we have developed.

Suppose we are given a measurement cover $\MM$. Recall that $\MM$ covers $X$, \ie $\bigcup \MM = X$.

We shall define a \emph{no-signalling empirical model} for $\MM$ to be a compatible family for the cover $\MM$ with respect to the presheaf $\DR\EE$. This means that for each measurement context $C \in \MM$, there is a distribution $e_C \in \DR\EE(C)$. Moreover, this family of distributions is compatible in the sense of the sheaf condition: for all $C, C' \in \MM$,
\[ e_C | C \cap C' = e_{C'} | C \cap C' . \]
In the case of Bell-type scenarios, this is readily seen to coincide with the usual notion of no-signalling. For example, in the bipartite case, consider  contexts $C = \{ m_a, m_b \}$, $C' = \{ m_a , m'_b \}$, with a choice of measurement each for Alice and Bob. Fix $s_0 \in \EE(\{m_a\})$, which assigns some outcome to $m_a$. Then the compatibility condition implies that
\[ \sum_{s \in \EE(C), s | m_a = s_0} e_C(s) \;\; = \;\; \sum_{s' \in \EE(C'), s' | m_a = s_0} e_{C'}(s') . \]
This says that the probability for Alice to get the outcome specified by $s_0$ for her measurement $m_a$ is the same, whether we marginalize over the possible outcomes for Bob when he makes the measurement $m_b$, or the measurement $m'_b$.  In other words, Bob's choice of measurement cannot influence Alice's outcome. This is exactly the standard definition of no-signalling.

We should also note, as a boundary case, that $\EE(\varnothing)$ is a one-element set, and $\DR\EE(\varnothing)$ is again a one-element set. Thus if contexts have empty intersection, the compatibility condition is trivially satisfied.

The general notion of compatible family for arbitrary covers $\MM$ applies to a much wider range of situations than Bell-type scenarios; later we will show that the empirical models which can be represented as quantum mechanical systems satisfy this more general form of no-signalling.

We shall only consider no-signalling models in this paper, so henceforth we shall simply speak of \emph{empirical models}.

\subsection{Examples}
\label{exsec}

We shall now show how some standard examples appear in our formalism.

%\subsubsection{A PR Box}

Consider a bipartite Bell-type scenario, where Alice has two possible measurements $\{ a, a' \}$, and Bob has $\{ b, b' \}$. There are two possible outcomes, $0$ or $1$, for each measurement.
%Thus this is a $(2,2,2)$ scenario.

Thus there are four maximal measurement contexts: 
\[ \{ a, b \}, \{ a', b \}, \{ a, b' \}, \{ a', b' \} \]
which index the rows of the following table:
\begin{center}
\begin{tabular}{l||c|c|c|c|}
& $(0, 0)$ & $(1, 0)$ & $(0, 1)$ & $(1, 1)$    \\  \hline \hline
$(a, b)$ & $p_1$ & $p_2$ & $p_3$ & $p_4$  \\  \hline
$(a', b)$ & $p_5$ & $p_6$ & $p_7$ & $p_8$  \\  \hline
$(a, b')$ & $p_9$ & $p_{10}$ & $p_{11}$ & $p_{12}$  \\  \hline
$(a', b')$ & $p_{13}$ & $p_{14}$ & $p_{15}$ & $p_{16}$  \\  \hline
\end{tabular}
\end{center}

\noindent The rows of this table correspond to the sets of sections $\EE(C)$, where $C$ ranges over the maximal measurement contexts. Thus, for example, the cell labelled with $p_2$ corresponds to the section $\{ a \mapsto 1, b \mapsto 0 \}$ in $\EE(C)$, where $C = \{ a, b \}$.

The table specifies a weight $p_i$ for each of these sections; in the standard case of probabilistic models, these will be non-negative reals, such that the values along each row sum to 1, and hence form a probability distribution. The  distributions $e_C$ for each maximal context $C$ collectively specify what we are calling an empirical model; and the no-signalling condition corresponds exactly to the compatibility condition on this family of distributions.

As a specific example, consider the following table:
\begin{center}
\begin{tabular}{ll|ccccc}
A & B & $(0, 0)$ & $(1, 0)$ & $(0, 1)$ & $(1, 1)$  &  \\ \hline
$a$ & $b$ & $1/2$ & $0$ & $0$ & $1/2$ & \\
$a'$ & $b$ & $3/8$ & $1/8$ & $1/8$ & $3/8$ & \\
$a$ & $b'$ & $3/8$ & $1/8$ & $1/8$ & $3/8$ &  \\
$a'$ & $b'$ & $1/8$ & $3/8$ & $3/8$ & $1/8$ & 
\end{tabular}
\end{center}
We shall use this model later to give a proof of Bell's theorem \cite{bell1964einstein}.

As another example, consider:
\begin{center}
\begin{tabular}{l|ccccc}
& $(0, 0)$ & $(1, 0)$ & $(0, 1)$ & $(1, 1)$  &  \\ \hline
$(a, b)$ & $1/2$ & $0$ & $0$ & $1/2$ & \\
$(a', b)$ & $1/2$ & $0$ & $0$ & $1/2$ & \\
$(a, b')$ & $1/2$ & $0$ & $0$ & $1/2$ & \\
$(a', b')$ & $0$ & $1/2$ & $1/2$ & $0$ & 
\end{tabular}
\end{center}
This is a PR box \cite{popescu1994quantum}.

We can also consider models over other semirings of weights. For example, the following is a specification of the possibilistic version of a  non-local Hardy model \cite{hardy1993nonlocality}, with weights in the boolean semiring.
It can be viewed as specifying the \emph{support} of a standard probabilistic Hardy model.

\begin{center}
\begin{tabular}{l|cccc} 
 &  $(0, 0)$ & $(1, 0)$ & $(0, 1)$ & $(1, 1)$ \\ \hline
$(a, b)$ &  $1$ &  $1$ &  $1$ &  $1$ \\
$(a', b)$ &   $0$ &  $1$ &  $1$ &  $1$ \\
$(a, b')$ &  $0$ &  $1$ &  $1$ & $1$ \\
$(a', b')$ &   $1$ &  $1$ &  $1$ & $0$ \\
\end{tabular}
\end{center}

\section{Global Sections}
\label{globsecsec}

We shall now show how the structures we have exposed in our mathematical description of empirical models can be brought to bear on the analysis of non-locality and contextuality.

We have already observed that the presheaf of events $\EE$ is in fact a sheaf; it is natural to ask if the same holds for the presheaf $\DR \EE$. Indeed, since empirical models are compatible families for this presheaf, to say that the sheaf condition holds for such a family $\{ e_C \}_{C \in \MM}$, with respect to a measurement cover $\MM$, is to say that there exists a  \emph{global section} $d \in \DR\EE(X)$, defined on the entire set of measurements $X$.
Such a global section defines a distribution on the set $\EE(X) = O^X$, which specifies assignments of outcomes to all measurements. Moreover, this distribution must restrict to yield the probabilities specified by the empirical model on all the  measurement contexts in $\MM$: \ie for all $C \in \MM$, 
$d | C = e_C$.
Thus the existence of a global section for the empirical model corresponds exactly to the existence of a distribution defined on all measurements, which marginalizes to yield the empirically observed probabilities. This places the idea of \emph{extendability} of probability distributions, as studied  in pioneering work by Fine \cite{fine1982hidden}, in a canonical and general mathematical form.

We can say more than this. A global assignment $s \in \EE(X) = O^X$, \ie a global section of the sheaf $\EE$, can be seen as a canonical form of \emph{deterministic hidden variable}, which assigns a definite outcome to each measurement, independent of the measurement context in which it appears. This yields an assignment $s | C$ for each measurement context $C$.
A global section $d  \in \DR\EE(X)$ specifies  a distribution on this canonical set of deterministic hidden variables. 
Each $s \in O^X$ induces a distribution $\delta_s \in \DR\EE(X)$, where $\delta_s(s) = 1$, and $\delta_s(s') = 0$ if $s \neq s'$. The distribution induced by $s$ on each measurement context $C$ is $\delta_s | C$;
note that $\delta_s | C = \delta_{s | C}$.
Now we have:
\[ \fl \qquad e_C (s) \; = \; d | C (s) \; = \; \sum_{s' \in \EE(X), s' | C = s} d(s') \; = \; \sum_{s' \in \EE(X)} \delta_{s' | C} (s) \cdot d(s') \; = \; \sum_{s' \in \EE(X)} \delta_{s'} | C (s) \cdot d(s') . \]
Thus the condition that $d |C = e_C$ for each measurement context $C$ says exactly that we reproduce the empirically observed probabilities $e_C(s)$ by \emph{averaging} over the hidden variables with respect to the distribution $d$. 

It is also easily verified that for each  context $C$, and $s' \in \EE(C)$:
\[ \delta_s | C(s') \;\; = \;\; \prod_{x \in C} \delta_{s | \{x\}} (s'|\{x\}) . \]
That is, the probability distribution determined by $s$ factors as a product of the probabilities assigned to the individual measurements, independent of the context in which they appear.  We shall define a general version of this  \emph{factorizability} property later, in Section~\ref{gsechvsec}.

If we specialize to the case of Bell-type scenarios, we see that factorizability corresponds to \emph{Bell locality} \cite{bell1964einstein}. For example, in a context $\{ a, b \}$, where $a$ is a measurement for Alice, and $b$ a measurement for Bob, then for a joint assignment of outcomes $\{ a \mapsto o_1, b \mapsto o_2 \}$, it says that the probability of this joint outcome determined by the hidden variable $s$ is the product of the probabilities it determines for $\{ a \mapsto o_1 \}$ and $\{ b \mapsto o_2 \}$ .
In other situations, it corresponds to a form of \emph{non-contextuality} at the level of distributions.

We can summarize this discussion as follows:

%\begin{center}

\begin{proposition}
\label{globsecdhvprop}
The existence of a global section for an empirical model implies the existence
 of a \emph{local (or non-contextual) deterministic hidden-variable model} which realizes it.\footnote{Note that `deterministic hidden variable model' means that the model is deterministic for each fixed value of the hidden variable.}
\end{proposition}
%\end{center}

We note also that, as we shall show later (see Theorem~\ref{equivth}), apparently more general forms of realization of empirical models by factorizable hidden-variable models, in which the hidden variables are not required to be deterministic, are in fact equivalent to canonical realizations by global sections. Thus the entire issue of non-locality and contextuality --- \ie the existence of empirical models which have no such realizations --- is equivalently formulated as the non-existence of global sections for the corresponding compatible families.

Thus we have a characterization of the phenomena of locality and non-contextuality in terms of \emph{obstructions to the existence of global sections}, a central issue in the pervasive applications of sheaves in geometry, topology, analysis and number theory. This opens the door to the use of the powerful methods of sheaf theory in the study of non-locality and contextuality.

\section{Existence of Global Sections}

The discussion in the previous section motivates the following problem:
\begin{center}
\fbox{Given an empirical model, determine if it has a global section.}
\end{center}
We shall give a general linear-algebraic method for answering this question, which as we have seen  is equivalent to the question of whether there exists a realization of the model by local or non-contextual hidden variables.

\subsection{The Incidence Matrix}

We are given a measurement cover $\MM$.
The first (and main) step is to construct a matrix $\MB$ of 0's and 1's, which we shall call the \emph{incidence matrix} of $\MM$. This matrix is defined using only $\MM$ and the event sheaf $\EE$, and can be applied to any empirical model expressed as a compatible family for $\MM$, with respect to any distribution functor $\DR$.

To define the incidence matrix, we firstly form the disjoint union $\coprod_{C \in \MM} \EE(C)$ of all the sections over the contexts in $\MM$, and then specify an enumeration
$s_1, \ldots , s_p$ of this set. We also specify an enumeration $t_1, \ldots , t_q$ of all the global assignments $t_j \in O^X$, \ie the global sections of the sheaf $\EE$. We then form the $(p \times q)$-matrix $\MB$, with entries defined as follows:
\[ \MB[i,j] = \left\{ \begin{array}{lr}
1, & t_j | C = s_i \; (s_i \in \EE(C)), \\
0, & \mbox{otherwise.}
\end{array}
\right.
\]
Conceptually, this matrix represents the tuple of restriction maps
\[ \EE(X) \lrarr \prod_{C \in \MM} \EE(C) :: s \mapsto (s | C)_{C \in \MM} . \]
To see this, note that for each $C$ we have the embedding 
\[ \EE(C) \lrarr \Pow(\EE(C)) :: s \mapsto \{ s \} . \]
Thus we obtain a map $\EE(X) \lrarr \prod_{C \in \MM} \Pow(\EE(C))$. Now we use the isomorphism
\[ \prod_{i \in I} \Pow(X_i) \; \cong \; \Pow(\coprod_{i \in I} X_i) \]
to obtain a map $\EE(X) \lrarr \Pow(\coprod_{C \in \MM} \EE(C))$. Such a map is the same thing as a relation
\[ R \;\; \subseteq \;\; \EE(X) \times \coprod_{C \in \MM} \EE(C) . \]
The incidence matrix is the boolean matrix representation of this relation.
Viewing it as a $0/1$-matrix over the semiring $R$,
it acts by matrix multiplication on distributions in $\DR\EE(X)$, represented as row vectors:
\[ d \mapsto (d | C)_{C \in \MM} . \]
Thus the image of this map will be the set of families $\{ e_C \}_{C \in \MM}$ which arise from global sections.

\subsection{Example: Bell-Type Scenarios}

We shall illustrate this construction for Bell-type scenarios. Following standard terminology, we shall refer to a Bell-type scenario with $n$ parts, each of which has $k$ possible measurements, each with $l$ possible outcomes, as of $(n,k,l)$-type.
Note that for a system of $(n,k,l)$-type, there are $k^n$  measurement contexts, for each of which there are $l^n$ possible assignments of outcomes. Thus there are $(kl)^n$ sections over the  contexts. The set of all measurements is of size $kn$, and there are $l^{kn}$ global assignments. Thus the incidence matrix in this case will be of size $(kl)^n \times l^{kn}$.
Each row of the matrix will contain $l^{(k-1)n}$ 1's.

We shall describe the (2,2,2) case explicitly. In this case, the matrix is $16 \times 16$.
We shall use the enumeration of sections over  contexts given in the table in Figure~1.
\begin{figure}
\label{enumfig}
\begin{center}
\begin{tabular}{l||c|c|c|c|}
& $(0, 0)$ & $(1, 0)$ & $(0, 1)$ & $(1, 1)$    \\  \hline \hline
$(a, b)$ & $s_1$ & $s_2$ & $s_3$ & $s_4$  \\  \hline
$(a', b)$ & $s_5$ & $s_6$ & $s_7$ & $s_8$  \\  \hline
$(a, b')$ & $s_9$ & $s_{10}$ & $s_{11}$ & $s_{12}$  \\  \hline
$(a', b')$ & $s_{13}$ & $s_{14}$ & $s_{15}$ & $s_{16}$  \\  \hline
\end{tabular}
\caption{Enumeration of sections}
\end{center}
\end{figure}
We shall also use an evident enumeration of global sections obtained by viewing them as binary strings, where the $i$'th bit indicates the assignment of an outcome to the $i$'th measurement, listed as $a, a', b, b'$.

The incidence matrix is then as follows.
{\small
\[  \left[ \begin{array}{cccccccccccccccc}
1 & 1 & 1 & 1 & 0 & 0 & 0 & 0 & 0 & 0 & 0 & 0 & 0 & 0 & 0 & 0 \\ 
0 & 0 & 0 & 0 & 1 & 1 & 1 & 1 & 0 & 0 & 0 & 0 & 0 & 0 & 0 & 0 \\ 
0 & 0 & 0 & 0 & 0 & 0 & 0 & 0 & 1 & 1 & 1 & 1 & 0 & 0 & 0 & 0 \\ 
0 & 0 & 0 & 0 & 0 & 0 & 0 & 0 & 0 & 0 & 0 & 0 & 1 & 1 & 1 & 1 \\ 
1 & 0 & 1 & 0 & 1 & 0 & 1 & 0 & 0 & 0 & 0 & 0 & 0 & 0 & 0 & 0 \\ 
0 & 1 & 0 & 1 & 0 & 1 & 0 & 1 & 0 & 0 & 0 & 0 & 0 & 0 & 0 & 0 \\ 
0 & 0 & 0 & 0 & 0 & 0 & 0 & 0 & 1 & 0 & 1 & 0 & 1 & 0 & 1 & 0 \\ 
0 & 0 & 0 & 0 & 0 & 0 & 0 & 0 & 0 & 1 & 0 & 1 & 0 & 1 & 0 & 1 \\ 
1 & 1 & 0 & 0 & 0 & 0 & 0 & 0 & 1 & 1 & 0 & 0 & 0 & 0 & 0 & 0 \\ 
0 & 0 & 0 & 0 & 1 & 1 & 0 & 0 & 0 & 0 & 0 & 0 & 1 & 1 & 0 & 0 \\ 
0 & 0 & 1 & 1 & 0 & 0 & 0 & 0 & 0 & 0 & 1 & 1 & 0 & 0 & 0 & 0 \\ 
0 & 0 & 0 & 0 & 0 & 0 & 1 & 1 & 0 & 0 & 0 & 0 & 0 & 0 & 1 & 1 \\ 
1 & 0 & 0 & 0 & 1 & 0 & 0 & 0 & 1 & 0 & 0 & 0 & 1 & 0 & 0 & 0 \\ 
0 & 1 & 0 & 0 & 0 & 1 & 0 & 0 & 0 & 1 & 0 & 0 & 0 & 1 & 0 & 0 \\ 
0 & 0 & 1 & 0 & 0 & 0 & 1 & 0 & 0 & 0 & 1 & 0 & 0 & 0 & 1 & 0 \\ 
0 & 0 & 0 & 1 & 0 & 0 & 0 & 1 & 0 & 0 & 0 & 1 & 0 & 0 & 0 & 1 \\
\end{array}
\right]
\]}
This matrix has rank 9. We shall give a general formula for the rank of the incidence matrix in Proposition~\ref{rankprop}, and apply it to the $(n,2,2)$ cases in Section~\ref{explicitformsubsec}.

We note that in the case of Bell-type scenarios, incidence matrices have been studied as `transfer matrices' in \cite{basoalto2003belltest}; see also \cite{pironio2005lifting}.\footnote{We thank one of the journal referees for bringing this connection to our attention.} Our account generalizes this to arbitrary measurement covers, and also provides a clear conceptual derivation of these matrices in terms of the restriction maps.
%\footnote{Verified using Mathematica\texttrademark.}

\subsection{Global Sections as Solutions of Linear Systems}

Now we consider an empirical model $\{ e_C \}$, defined with respect to the distribution functor $\DR$. Such a model assigns a weight in the semiring $R$ to each section $s_i \in \EE(C)$.
Thus it can be specified by a vector $\vv$ of length $p$, where $\vv[i] = e_C(s_i)$.
We can also introduce a vector $\xx$ of length $q$ of `unknowns', one for each global section $t_j \in \EE(X)$.
Now a solution for the linear system $\MB \xx = \vv$ will be a vector of values in $R$, one for each $t_j$. To ensure that this vector is a distribution, we augment $\MB$ with an extra row, every entry in which is 1, and similarly augment $\vv$ with an extra element, also set to $1$. A solution for this augmented system will enforce the constraint
\[ \xx[1] + \cdots + \xx[q] = 1 \]
and hence ensure that the assignment of weights defines a distribution on $\EE(X)$.
The remaining equations ensure that this distribution restricts to yield the weight specified by the empirical model for each section $s_j$.

\begin{proposition}
Let $\MB'$ be the augmented incidence matrix, and $\vv'$ the augmented vector corresponding to an empirical model $e$ over the distribution functor $\DR$. 
Solutions to this system of equations $\MB' \xx = \vv'$ in $R$ correspond bijectively to global sections for $e$.
\end{proposition}

We also note that in the case of Bell-type scenarios of $(n,k,l)$-type, it is not necessary to use the augmented system; solutions of the equation $\MB \xx = \vv$ will automatically be distributions. This follows easily from the regular structure of the incidence matrices for these cases.

\subsection{Examples}

We shall consider a number of examples, based on the models of $(2,2,2)$-type discussed in Section~\ref{exsec}.

\subsubsection{The Bell Model}

We look again at the Bell model
\begin{center}
\begin{tabular}{l|ccccc}
& $(0, 0)$ & $(1, 0)$ & $(0, 1)$ & $(1, 1)$  &  \\ \hline
$(a, b)$ & $1/2$ & $0$ & $0$ & $1/2$ & \\
$(a', b)$ & $3/8$ & $1/8$ & $1/8$ & $3/8$ & \\
$(a, b')$ & $3/8$ & $1/8$ & $1/8$ & $3/8$ &  \\
$(a', b')$ & $1/8$ & $3/8$ & $3/8$ & $1/8$ & 
\end{tabular}
\end{center}

We are interested in finding a solution in the non-negative reals, \ie a probability distribution on the global assignments $\EE(X)$. This amounts to solving the linear system over the reals, subject to the constraint $\xx \geq \mathbf{0}$; \ie to a linear programming problem.
It is easy in this case to give a direct argument that there is no such solution, and hence that the above model has no hidden-variable realization, thus proving Bell's theorem \cite{bell1964einstein}.

\begin{proposition}
The Bell model has no global section.
\end{proposition}
\begin{proof}
We focus on 4 out of the 16 equations, corresponding to rows $1$, $6$, $11$ and $13$ of the incidence matrix. We write $X_i$ rather than $\xx[i]$.
\[
\begin{array}{lclclclcc}
X_1 & + & X_{2} & + & X_{3} & + & X_{4} & = & 1/2 \\
X_2 & + & X_{4} & + & X_{6} & + & X_{8} & = & 1/8 \\
X_3 & + & X_{4} & + & X_{11} & + & X_{12} & = & 1/8 \\
X_1 & + & X_{5} & + & X_{9} & + & X_{13} & = & 1/8 \\
\end{array}
\]
Adding the last three equations yields
\[ X_1 + X_2 + X_3 + 2X_4 + X_5 + X_6 + X_8 + X_9  + X_{11} + X_{12} + X_{13}  \; = \; 3/8 .\]
Since all these terms must be non-negative, the left-hand side of this equation must be greater than or equal to the left-hand side of the first equation, yielding the required contradiction.
\end{proof}

\subsubsection{The Hardy Model}

Now we consider the possibilistic version of the Hardy model, specified by the following table.
\begin{center}
\begin{tabular}{l|cccc} 
 &  $(0, 0)$ & $(1, 0)$ & $(0, 1)$ & $(1, 1)$ \\ \hline
$(a, b)$ &  $1$ &  $1$ &  $1$ &  $1$ \\
$(a', b)$ &   $0$ &  $1$ &  $1$ &  $1$ \\
$(a, b')$ &  $0$ &  $1$ &  $1$ & $1$ \\
$(a', b')$ &   $1$ &  $1$ &  $1$ & $0$ \\
\end{tabular}
\end{center}
This is obtained from a standard probabilistic Hardy model by replacing all positive entries by $1$; thus it can be interpreted as the \emph{support} of the probabilistic model.

In this case, we are interested in the existence of a solution over the boolean semiring.
This corresponds to a boolean satisfiability problem.  For example, the equation specified by the first row of the incidence matrix corresponds to the clause
\[ X_1 \OR X_2 \OR X_3 \OR X_4 \]
while the fifth yields the formula
\[ \neg X_1 \AND \neg X_3 \AND \neg X_ 5 \AND \neg X_7 . \]
A solution is an assignment of boolean values to the variables which simultaneously satisfies all these formulas.
Again, it is easy to see by a direct argument that no such assignment exists.

\begin{proposition}
The possibilistic Hardy model has no global section over the booleans.
\end{proposition}
\begin{proof}
We focus on the four formulas corresponding to rows $1$, $5$, $9$ and $16$ of the incidence matrix:
\[ \begin{array}{ccccccc}
X_1 & \OR & X_2 & \OR & X_3 & \OR & X_4 \\
 \neg X_1 & \AND & \neg X_3 & \AND & \neg X_ 5 & \AND & \neg X_7 \\
  \neg X_1 & \AND & \neg X_2 & \AND & \neg X_ {9} & \AND & \neg X_{10} \\
   \neg X_4 & \AND & \neg X_8 & \AND & \neg X_ {12} & \AND & \neg X_{16} 
\end{array}
\]
Since every disjunct in the first formula appears as a negated conjunct in one of the other three formulas, there is no satisfying assignment.
\end{proof}

To understand the significance of this result, we note the following general fact.

\begin{proposition}
\label{vvbprop}
Let $\MB$ be the incidence matrix for a cover $\MM$, and let $\vv$ be the vector of non-negative reals corresponding to a probabilistic model over $\MM$. Let $\vv_b$ be the boolean vector obtained by replacing each non-zero element of $\vv$ by $1$. If the system $\MB \xx = \vv$ has a solution over the non-negative reals, then the system $\MB \xx = \vv_b$ has a solution over the booleans.
\end{proposition}
\begin{proof}
This follows simply from the fact that the map  from the non-negative reals to the booleans which takes all non-zero elements to 1 is a semiring homomorphism.
\end{proof}

It follows that, if the support of a probabilistic model has no global section with respect to the boolean distribution functor $\DB$, then the probabilistic model itself has no global section with respect to the probability distribution functor $\DP$.
Thus the argument given above implies that the probabilistic Hardy model also has no global section, and hence is non-local.

The converse to Proposition~\ref{vvbprop} is false. Indeed, the Bell model, which as we have seen has no probabilistic global section, \textit{does} have a boolean global section for its support.
This is easy to show directly, but also follows from the general results in \cite{MF2011}, which show that Hardy models are complete for the (2,2,2)-type cases, and in particular that there must be at least three sections excluded from the support in order for non-locality to hold, while the Bell model has only two zero entries.

In this sense, we can say that the Hardy model satisfies a \textit{stronger} non-locality property than the Bell model.
In general, we say that a probabilistic model is \emph{probabilistically non-extendable} it it has no global section over $\DP$, and \emph{possibilistically non-extendable} if its support has no global section over $\DB$. We have seen that possibilistic  non-extendability is strictly stronger than probabilistic non-extendability.

\section{Negative Probabilities}

We shall now consider the question of extendability over real-valued distributions $\DReal$, i.e. signed probability measures.  Formally, this simply amounts to solving the linear system over the reals, with no additional constraints. Conceptually, this allows the introduction of \emph{negative probabilities} in the extended model.  Of course, these marginalize to yield standard non-negative probabilities in the measurement contexts stipulated by the empirical model. Thus the usual relative-frequency interpretation of the actually observed statistics is maintained.

The appearance of negative probabilities in quantum mechanics has a long history, which we shall sketch in the Postlude (Section 10).  In this section, we shall prove that all empirical models are extendable with respect to signed probability measures.  In fact, there is an \textit{equivalence} between extendability under signed measures and no-signalling. Note that the class of no-signalling models is strictly larger than the  empirical models of this type that arise in quantum mechanics.  For example,  it includes the superquantum Popescu-Rohrlich boxes \cite{popescu1994quantum}.

The result therefore shows that negative probabilities, in themselves, cannot characterize quantum mechanics. This runs contrary to an impression which might be gained from the literature.
For example, Feynman writes: ``The only difference between a probabilistic classical world and the equations of the quantum world is that somehow or other it appears as if the probabilities would have to go negative \ldots '' \cite[p.\,480]{feynman1982simulating}.   In fact, the introduction of negative probabilities yields the entire no-signalling world.

\subsection{Solving the Linear System Over the Reals}
Given an empirical model $e$ over $\DReal$, our aim is to find a global section.
The existence of such a global section for $e$, which is represented by the real vector $\vv$, reduces to the existence of a solution for  the linear system $\MB \xx = \vv$ over the reals, with no additional constraints.

Note that there is no semiring homomorphism from the \textit{reals} to the booleans.
Indeed, if there were such a homomorphism $h$, we would have:
\[ 0 = h(0) = h(1 + (-1)) = h(1) \vee h(-1) = 1 \vee h(-1) = 1. \]
A similar argument shows that there is no homomorphism from the reals to the non-negative reals.
Thus there is no result analogous to Proposition~\ref{vvbprop}, and it is possible for the linear system to be solvable over the reals, while no solution exists over the non-negative reals or the booleans.

We shall now show that such solutions exist for all no-signalling probabilistic empirical models, over arbitrary measurement covers. This substantially generalizes previous results, e.g. Theorem~1 in \cite{pironio2005lifting}. 
%Our proof will make use of some of the ideas in \cite{pironio2005lifting}, adapted to the more general setting.

We introduce some notation. We fix a standard set of outcomes
$O \, := \, \{ 1, \ldots , l \}$.
Given a cover $\MM$, we define the set of  \emph{partial contexts}:
\[ \UU \, := \, \{ U \subseteq X \mid \exists C \in \MM. \, U \subseteq C \} . \]
For each $U \in \UU$ and $p \geq 0$, we define 
\[ \EEj{p}(U) \, := \, \{ s \in \EE(U) \mid \card{s^{-1}(\{ 1 \})} \leq p \} , \]
the set of sections which map at most $p$ measurements to the outcome $1$.
%$\Eno(U) := \{ 2, \ldots , l \}^U$, the set of sections with values excluding $1$.

Given a section $s$, we write $\sm{j}$ for the section defined by: 
\[ \sm{j}(m) = j, \qquad \sm{j}(m') = s(m'), \;\; (m' \neq m) . \]
Finally, given an empirical model $e$, we define
\[ \ej{p} := \{ e_U(s) \mid U \in \UU, s \in \EEj{p}(U) \} . \]

\begin{proposition}
\label{margdetprop}
Let $e$ be a probabilistic model over $\MM$. Then $e$ is linearly determined by $\ej{0}$.\end{proposition}
\begin{proof}
We shall prove that we can infer $\ej{p+1}$ from $\ej{p}$; the fact that $e$ is determined by $\ej{0}$ then follows by induction.

Consider $U \in \UU$, $s \in \EEj{p}$, and $m \in U$. Let $U' := U \setminus \{ m \}$. Using compatibility,
\[ e_{U'}(s | U')  \; = \; \sum_j e_U(\sm{j}) . \]
Hence 
\begin{equation}
\label{margineq}
e_U(\sm{1}) \; = \; e_{U'}(s | U') \, - \,  \sum_{j \neq 1} e_U(\sm{j}) . 
\end{equation}
All the terms on the RHS of this equation are in $\ej{p}$; while every element of $\ej{p+1}$ can be written in the form of the LHS.

Unwinding the induction, every number $e_C(s)$ is given by a linear combination of values in $\ej{0}$, obtained by back-substitution in (\ref{margineq}).
\end{proof}

We can consider probabilistic models as real vectors, as in the previous section. For a given  cover $\MM$, the dimension of the ambient vector space will be $t := \sum_{C \in \MM} l^{\card{C}}$.

\begin{proposition}
\label{nosigdimprop}
The dimension of the subspace of $\Rt$ spanned by the vectors arising from probabilistic empirical models is bounded above by $D := \sum_{U \in \UU} (l-1)^{\card{U}}$.
\end{proposition}
\begin{proof}
From Proposition~\ref{margdetprop}, we know that any probabilistic empirical model is determined by a vector of $D$ numbers. Moreover, the corresponding map $L : \Real^D \rarr \Rt$ defined by the equations (\ref{margineq})  is linear. Hence the subspace spanned by the probabilistic empirical models is contained in the image of $L$, and has dimension $\leq D$.
\end{proof}

This gives us an upper bound on the dimension of the  linear subspace generated by the `no-signalling polytope' over the measurement cover $\MM$.
We shall now give a lower bound on the dimension of the linear space spanned by the non-contextual models over $\MM$ --- \ie those arising from global sections.

Given a cover $\MM$,  we have the set $\UU$ of partial contexts.
Given $U \in \UU$ and $s \in \EEj{0}(U)$, we can define the global assignment $\tus : X \rarr O$:
\[ \tus(m) = s(m), \quad (m \in U), \qquad \tus(m) = 1, \quad (m \not\in U) . \]
Note that $\tus = \tusp$ implies $U = U'$ and $s = s'$.
Each such assignment $\tus$ defines a column vector $\vus = \MB[\_ \, , \tus]$.
There are clearly  $D = \sum_{U \in \UU} (l-1)^{\card{U}}$ such assignments.

\begin{proposition}
\label{locdimprop}
The set of vectors $\{ \vus \}_{U \in \UU, s \in \EEj{0}(U)}$
 is linearly independent. Thus the dimension of the linear subspace of $\Rt$ spanned by the vectors arising from global sections is bounded below by $D$.
\end{proposition}
\begin{proof}
Suppose that $\sum_{U, s} \mus \vus = \mathbf{0}$. 
We shall show that $\mus = 0$ for all $U, s$, by complete induction on $\card{X \setminus U}$. 

Given some $U', s'$, we choose a row of the incidence matrix $(C, s_0)$ such that $U' \subseteq C$ and $\tusp | C = s_0$, so that $s_0 | U' = s'$.
Note that, for any $U'', s''$,  $\MB[(C, s_0), \tusq] = 1$ if and only if $\tusq | C = s_0$, if and only if $U'' \cap C = U'$, and $s'' | U' = s'$.
If $\tusq \neq \tusp$, we must then have $U'' \supset U'$; so by induction hypothesis, $\musq = 0$.
Using the $(C, s_0)$ component of the vector equation $\sum_{U, s} \mus \vus = \mathbf{0}$, we conclude that $\musp = 0$.
\end{proof}

We now come to our main result.

\begin{theorem}
\label{dimthm}
Let $\MM$ be any measurement cover.  The linear subspaces generated by the non-contextual and the no-signalling models over this cover coincide, with dimension $D$.
\end{theorem}
\begin{proof}
Since the non-contextual models are a subset of the no-signalling models, this follows immediately from the matching lower bound on the dimension of the local subspace from Proposition~\ref{locdimprop}, and upper bound on the dimension of the no-signalling space from Proposition~\ref{nosigdimprop}.
\end{proof}

As an immediate consequence of this result, we have:

\begin{theorem}
\label{lineqthm}
Let $\MM$ be any measurement cover, and $e$ a probabilistic model over this cover, with corresponding vector $\vv \in \Rt$.
Then the linear system $\MB \xx = \vv$ has a solution over the reals.
\end{theorem}

We can also apply this result to the incidence matrix. Let $\MB$ be the incidence matrix defined over a cover $\MM$ and set of outcomes $O$.
\begin{proposition}
\label{rankprop}
The rank of $\MB$, as a matrix over the reals, is $D$.
\end{proposition}
\begin{proof}
The incidence matrix defines a linear map from the vector space generated by the global assignments $O^X$ into $\Rt$. By Theorem~\ref{dimthm}, the dimension of the image of this map is $D$.
\end{proof}

\subsection{Example: The PR Box}

We consider the PR box:
\begin{center}
\begin{tabular}{l|ccccc}
& $(0, 0)$ & $(1, 0)$ & $(0, 1)$ & $(1, 1)$  &  \\ \hline
$(a, b)$ & $1/2$ & $0$ & $0$ & $1/2$ & \\
$(a', b)$ & $1/2$ & $0$ & $0$ & $1/2$ & \\
$(a, b')$ & $1/2$ & $0$ & $0$ & $1/2$ & \\
$(a', b')$ & $0$ & $1/2$ & $1/2$ & $0$ & 
\end{tabular}
\end{center}
A simple solution of the linear system for the PR box is  the vector
\[ [1/2, 0, 0, 0, - 1/2, 0, 1/2, 0, - 1/2, 1/2, 0, 0, 1/2, 0, 0, 0 ] . \]
This vector can be taken as giving a local hidden-variable realization of the PR box using negative probabilities. Similar realizations can be given for the other PR boxes.

\subsection{An Explicit Formula For The Dimension}
\label{explicitformsubsec}

We now consider some  symmetry properties of a cover $\MM$, and the associated family of partial contexts $\UU$. We define $\UUj$ to be the set of elements of $\UU$ of cardinality $j$, $0 \leq j \leq n$.
We say that $\MM$ is \emph{homogeneous} if the following conditions hold:

\begin{enumerate}
\item All the contexts $C \in \MM$ have the same number $n$ of elements.
\item Every set $U \in \UUj$ is a subset of the same number $N_j$ of contexts $C \in \MM$. Note that we must always have $N_0 = p$, where $p = \card{\MM}$.
%\item Each measurement appears in two contexts.
\end{enumerate}

We consider some examples:
\begin{itemize}
\item Every $(n, k, l)$-type Bell scenario is homogeneous, with $p = k^n$, and $N_j = k^{n-j}$.

\item The measurement cover described in Section~\ref{KSscensubsec}, corresponding to the Kochen-Specker proof from \cite{cabello1996bell}, is homogeneous, with $p = 9$, $n=4$, $N_1 = 2$, and $N_j = 1$ for all $2 \leq j \leq 4$.

\item Many of the constructions used in Kochen-Specker proofs are homogeneous, for example the cover corresponding to the Peres-Mermin magic square \cite{peres1990incompatible,mermin1990simple}, which consists of the rows and columns of the table
\begin{center}
\begin{tabular}{|c|c|c|}
\hline
$A$ & $B$ & $C$ \\ \hline
$D$ & $E$ & $F$ \\ \hline
$G$ & $H$ & $I$ \\ \hline
\end{tabular}
\end{center}
In this case, $p = 6$, $n = 3$, $N_1 = 2$, and $N_2 = N_3 = 1$.
\end{itemize}

\begin{proposition}
Let $\MM$ be a homogeneous cover. Then:
\[ D \; = \; \sum_{j=0}^n \frac{{n \choose j} p  (l-1)^j}{N_j}   . \]
\end{proposition}
\begin{proof}
From the definitions, 
\[ D \; = \; \sum_{U \in \UU} (l-1)^{\card{U}} \; = \;  \sum_{j=0}^n \card{\UUj} (l-1)^j  . \]
Homogeneity implies that $\card{\UUj} = p {n \choose j}/N_j$.
\end{proof}

We now apply this result to our examples:

\begin{itemize}
\item For $(n, k, l)$-type scenarios, we have
\[ D \; = \; \sum_{j=0}^n \frac{{n \choose j} k^n  (l-1)^j}{k^{n-j}}  \; = \; \sum_{j=0}^n {n \choose j} k^j  (l-1)^j  . \]
Applying the binomial theorem, we obtain $D = (k \cdot (l-1) + 1)^n$.
This retrieves the dimension established in \cite{pironio2005lifting}, with the minor difference that the value given 
%in \cite{pironio2005lifting} 
there is $D - 1$.
This apparent discrepancy arises because marginalization over the empty set is excluded in \cite{pironio2005lifting}, using the fact that, by normalization, $e_{\vn}(\vn) = 1$. However, in this case the equations (\ref{margineq}) are \textit{affine} rather than linear.

\item For the 18-vector cover, taking $l=2$, we obtain $D = 118$.
This can be compared with the dimension of the ambient vector space, which is $9 \cdot 2^4 = 144$. 

\item The corresponding value for the Peres-Mermin square is $D = 34$, with ambient dimension 48.
\end{itemize}

We can also apply this formula to the rank of the incidence matrix. For example, for $(n, 2, 2)$-scenarios, where the incidence matrix is of size $4^{n} \times 4^{n}$, the rank is $3^n$.
%It seems likely that interesting combinatorial formulas can be found for  other classes of covers.

This formula for the rank can be made visually apparent in this case, by noting that, with a suitable choice of enumeration for the rows and columns, the incidence matrices $\MB(n)$ have a self-similar inductive structure:
\[ \MB(1): \left[ \begin{array}{cccc}
1 & 1 & 0 & 0   \\ 
0 & 0 & 1 & 1  \\ 
1 & 0 & 1 & 0 \\ 
0 & 1 & 0 & 1  \\ 
\end{array}
\right]
\qquad 
\MB(n+1): \left[ \begin{array}{cccc}
\MB(n) & \MB(n) & 0 & 0   \\ 
0 & 0 & \MB(n) & \MB(n)  \\ 
\MB(n) & 0 & \MB(n) & 0 \\ 
0 & \MB(n) & 0 & \MB(n)  \\ 
\end{array}
\right]
\]

\subsection{Global Sections and No-Signalling}

No-signalling has been built into our notion of empirical model through the requirement of compatibility of the family $\{ e_C \}$. 
Note, though, that any family, whether compatible or not, gives rise to a linear system of equations
$\MB \xx = \vv$.
If this system has a solution, so that the family has a global section, it  is
\textit{automatically} compatible, and hence satisfies no-signalling.

\begin{proposition}
Let $d \in \DR\EE(X)$ be a global section. Then the family $\{ d | C \}_{C \in \MM}$ is compatible.
\end{proposition}
\begin{proof}
This follows immediately from the functoriality of restriction. For any $C, C' \in \MM$:
\[    \DR(\rmap{C}{C \cap C'}) \circ  \DR(\rmap{X}{C})(d) \; = \; \DR(\rmap{C}{C \cap C'}  \circ  \rmap{X}{C} )(d) \; = \; \DR(\rmap{X}{C \cap C'}) (d) \]
and thus $(d | C) | C \cap C' \; = \; d | C \cap C'$.
Similarly, $(d | C') | C \cap C' = d | C \cap C' $. Hence $d | C$ and $d | C'$ agree on their overlap.
\end{proof}

Combining this result with Theorem~\ref{lineqthm}, we obtain the following Theorem.

\begin{theorem}
\label{nsprop}
Probability models have local hidden-variable realizations with negative probabilities if and only if they satisfy no-signalling.
\end{theorem}

Thus we have a striking equivalence between no-signalling models, and those admitting local hidden-variable realizations with negative probabilities.

\section{Strong Contextuality}
\label{scsec}

Consider a probability model over a cover $\MM$. By Proposition~\ref{vvbprop}, if the model is extendable over $\DP$, its support is extendable over the booleans. This means that there is a boolean distribution $d$ on $\EE(X)$ which restricts to $\supp(e_C)$ for every context $C \in \MM$.
Such a distribution is simply a non-empty subset $S$ of $\EE(X)$. The condition that $d | C = \supp(e_C)$ means that, for all $s \in S$,  $s | C \in \supp(e_C)$ for every $C \in \MM$; and moreover, every section in $\supp(e_C)$ is of the form $s|C$ for some $s$ in $S$.

Given an empirical model $e$, we define the set  
\[ S_e \; := \; \{ s \in \EE(X) : \forall C \in \MM . \, s | C \in  \supp(e_C) \} . \]
Thus a consequence of the extendability of $e$ is that $S_e$ is non-empty.

We say that the model $e$ is \emph{strongly contextual} if this set $S_e$ is empty.
Whereas a global section for an empirical model $e$ completely determines its behaviour, asking for some assignment $s \in \EE(X)$ which is \textit{consistent} with the support of  $e$ is much weaker.
The negative property that \textit{not even one such assignment exists} is correspondingly much stronger.
Indeed, the Hardy model, which as we saw in the previous section is possibilistically non-extendable, is \textit{not} strongly contextual. The global assignment
\[ \{ a \mapsto 1, \; a' \mapsto 0, \; b \mapsto 1, \; b' \mapsto 0 \} \]
is in $S_e$ for this model. The Bell model similarly fails to be strongly contextual.

The question now arises: are there models coming from quantum mechanics which are strongly contextual in this sense?

We shall now show that the well-known GHZ models \cite{greenberger1989going}, of type $(n, 2, 2)$ for all $n > 2$, are strongly contextual. This will establish a strict hierarchy
\[ \mbox{Bell} < \mbox{Hardy} < \mbox{GHZ} \]
of increasing strengths of obstructions to non-contextual behaviour for these salient models.

The GHZ model of type $(n, 2, 2)$ can be specified as follows.
We label the two measurements at each part as $\mxi$ and $\myi$, and the outcomes as $0$ and $1$.
For each context $C$, every $s$ in the support of the model satisfies the following conditions:
\begin{itemize}
\item If the number of $Y$ measurements in $C$ is a multiple of 4, the number of $1$'s in the outcomes specified by $s$ is even.
\item If the number of $Y$ measurements  is $4k+2$, the number of $1$'s in the outcomes is odd.\end{itemize}
We will see later how a model with these properties can be realized in quantum mechanics.

\begin{proposition}
The GHZ models are strongly contextual, for all $n \geq 3$.
\end{proposition}
\begin{proof}
We consider the case where $n = 4k$, $k \geq 1$.
Assume for a contradiction that we have a global section $s \in S_e$ for the GHZ model $e$.

If we take $Y$ measurements at every part, the number of $1$ outcomes under the assignment is even.
 Replacing any two $Y$'s by $X$'s changes the residue class mod $4$ of the number of $Y$'s, and hence must result in the opposite parity for the number of $1$ outcomes under the assignment.
Thus for any $\myi$, $\myj$ assigned the \textit{same} value, if we substitute $X$'s in those positions they must receive \textit{different} values under $s$. Similarly, for any  $\myi$, $\myj$ assigned different values, the corresponding  $\mxi$, $\mxj$ must receive the same value.

Suppose firstly that not all $\myi$ are assigned the same value by $s$.
Then for some $i$, $j$, $k$, $\myi$ is assigned the same value as $\myj$, and $\myj$ is assigned a different value to $\myk$. Thus $\myi$ is also assigned a different value to $\myk$. Then $\mxi$ is assigned the same value as $\mxk$, and $\mxj$ is assigned the same value as $\mxk$. By transitivity, $\mxi$ is assigned the same value as $\mxj$, yielding a contradiction.

The remaining cases are where all $Y$'s receive the same value. Then any pair of $X$'s must receive different values. But taking any 3 $X$'s, this yields a contradiction, since there are only two values, so some pair must receive the same value.

The case when $n = 4k+2$, $k \geq 1$,  is proved in the same fashion, interchanging the parities.
When $n \geq 5$ is odd, we start with a context containing one $X$, and again proceed similarly.

The most familiar case, for $n=3$, does not admit this argument, which relies on having at least 4 $Y$'s in the initial configuration.  However, for this case one can easily adapt the well-known argument of Mermin using `instruction sets' \cite{mermin1990quantum} to prove strong contextuality. This uses a case analysis to show that there are 8 possible global sections satisfying the parity constraint on the 3 measurement combinations with 2 $Y$'s and 1 $X$; and all of these violate the constraint for the $XXX$ measurement.
\end{proof}

We shall also mention an elegant result due to Ray Lal (private communication).

\begin{proposition}[Lal]
The only strongly contextual no-signalling models of type $(2,2,2)$ are the PR boxes.
\end{proposition}

Thus strong contextuality actually \textit{characterizes} the PR boxes.

\subsection{Strong Contextuality and Maximal Non-Locality}

The property of strong contextuality is defined in a simple `qualitative' fashion, in terms of the support of a model. As we shall now see, for probabilistic models it is equivalent to a notion which can be defined in quantitative terms, and has been studied in this form in the special case of Bell-type scenarios.\footnote{We thank one of the journal referees for pointing out this connection.}

Suppose that $\{ e_C \}_{C \in \MM}$ is a model over the presheaf $\DP$. We consider convex decompositions
\begin{equation}
\label{condeceq}
e = \lambda L + (1-\lambda) q, \qquad 0 \leq \lambda \leq 1,
\end{equation}
where $L$ is a local model, and $q$ a no-signalling model. This means that, for every $C \in \MM$, and $s \in \EE(C)$, we have:
\[ e_C(s) = \lambda L_C(s) + (1-\lambda) q_C(s) . \]

We define the \emph{non-contextual fraction} of $e$ to be the supremum over all $\lambda$ appearing in such convex decompositions (\ref{condeceq}). This notion was introduced  for Bell-type scenarios in \cite{elitzur1992quantum}; see also \cite{barrett2006maximally,aolita2011fully}, where the terminology \emph{local fraction} is used. 
A model with local fraction $0$ is defined to be \emph{maximally non-local}. Geometrically, this corresponds to the model being on a face of the no-signalling polytope with no local elements.

In the general setting of models defined on arbitrary measurement covers, we say that a model $e$ is \emph{maximally contextual} if the non-contextual fraction of $e$ is $0$.

\begin{proposition}
\label{sceqmcprop}
A model $e$ is strongly contextual if and only if it it is maximally contextual.
\end{proposition}
\begin{proof}
Suppose firstly that $e$ admits a convex decomposition 
 (\ref{condeceq}). 
By the results of Section~\ref{globsecsec}, and also Theorem~\ref{equivth}, we can take $L$ to be a convex sum of deterministic models $\sum_i \mu_i \delta_{s_i}$, where each $s_i \in \EE(X)$ is a global assignment. If $\lambda > 0$, then from (\ref{condeceq}),  $s_i \in S_e$ for each $i$.
Thus strong contextuality implies maximal contextuality.

For the converse, suppose that $s \in S_e$.
Taking $L = \lambda \cdot \delta_s$, we shall define $q$ such that (\ref{condeceq}) holds. For each $C \in \MM$ and $s' \in \EE(C)$:
\[ q_C(s') \, := \, \frac{e_C(s') - \lambda \cdot \delta_{s | C}(s')}{1 - \lambda} . \]
It is easily verified that, for each $C$, $\sum_{s' \in \EE(C)} q_C(s') = 1$. To ensure that $q$ is always non-negative, we must have $\lambda \leq \inf_{C \in \MM} e_C(s | C)$.
Since this is the infimum of a finite set of positive numbers, we can find $\lambda > 0$ satisfying this condition.

It remains to verify that $q$ is no-signalling, \ie that $\{ q_C \}$ forms a compatible family.
Given $C, C' \in \MM$, fix $s_0 \in \EE(C \cap C')$. Now
\[ q_C | C \cap C'(s_0) = 1/(1 - \lambda) [(\sum_{s' \in \EE(C), s' | C \cap C' = s_0} e_C(s')) \;  - \; \lambda \cdot \delta_{s | C \cap C'}(s_0)] . \]
A similar analysis applies to $q_{C' }| C \cap C'(s_0)$. Using the compatibility of $e$, we conclude that $q_C | C \cap C' = q_{C'} | C \cap C'$.
\end{proof}

We can use this equivalence to give a characterization of maximal contextuality, and in particular of maximal non-locality, in terms of a \emph{constraint satisfaction problem}. In the case of dichotomic measurements, this reduces to a \emph{boolean satisfiability problem}.

We recall that a constraint satisfaction problem (CSP) \cite{montanari1974networks,kumar1992algorithms} is specified by a triple $(V, K, \RR)$, where $V$ is a finite set of  variables, $K$ is a finite set of values, and $\RR$ is a finite set of constraints. A constraint is a pair $(C, S)$, where $C \subseteq V$, and $S \subseteq K^C$. (It is more common to define a constraint as an ordered list of $k$ variables, and a set of $k$-tuples of values, but this is obviously equivalent to the version given.)
An assignment $s : V \rarr K$ satisfies a constraint $(C, S)$ if $s | C \in S$. A solution of the CSP $(V, K, \RR)$ is an assignment $s : V \rarr K$ which  satisfies every constraint in $\RR$.

Let $e$ be a model over a cover $\MM$, with outcome set $O$.
For each $C \in \MM$, we have the set $\SP_e(C) := \supp(e_C) \subseteq O^C$.
We can associate the CSP $(X, O, \{ \SP_e(C) \mid C \in \MM \})$ with $e$.

\begin{proposition}
A probabilistic model $e$ is maximally contextual if and only if the corresponding CSP has no solution. In particular, for Bell-type scenarios, $e$ is maximally non-local if and only if the corresponding CSP has no solution.
\end{proposition}
\begin{proof}
This follows directly from Proposition~\ref{sceqmcprop}, since global sections in the support of $e$ are clearly in bijective correspondence with solutions of the associated CSP.
\end{proof}

In the case of dichotomic measurements, the CSP reduces to a boolean satisfiability problem.
In this case, we interpret the two possible outcomes as truth values, and $X$ as a set of propositional variables.

Given a model $e$, we define the formula
\[ \phi_e \; := \;   \bigwedge_{C \in \MM} \; \bigvee_{s \in \SP_e(C)} \psi_s \]
where for a section $s \in O^C$ we define the corresponding formula
\[ \psi_s \; := \; \bigwedge_{m \in C, s(m) = \true} m \;\; \AND \;\; \bigwedge_{m \in C, s(m) = \false} \neg m . \]

\begin{proposition}
A probabilistic model $e$ with dichotomic measurements is maximally contextual if and only if the corresponding formula $\phi_e$ is unsatisfiable. In particular, for Bell-type scenarios, $e$ is maximally non-local if and only $\phi_e$ is unsatisfiable.
\end{proposition}

\section{Generic Strong Contextuality and Kochen-Specker Theorems}
\label{abskssec}

Let $e$ and $e'$ be models, such that the support of $e$ is included in the support of $e'$.
Then $S_e$ is included in $S_{e'}$; hence  if $e'$ is strongly contextual, so is $e$.
Thus by showing strong contextuality for a single model, we can show it for a whole class of models.

We shall fix our set of outcomes as $\{ 0, 1 \}$. This means that we can define subsets  of $\EE(C)$ by formulas $\phi_C$, with the elements of $C$ used as propositional variables. 
A section $s : C \rarr \{ 0, 1 \}$ can be viewed as a boolean assignment for these variables, and $\phi_C$ defines the set of its satisfying assignments.

We are interested in particular in the formula 
\[ \XOR(C) \; := \; \bigvee_{m \in C} (m \; \wedge \; \bigwedge_{m' \in C \setminus \{ m \}} \neg m') . \]
This is satisfied by those assignments with exactly one outcome set to $1$.

A Kochen-Specker-type result \cite{kochen1975problem} can be factored into two parts:
\begin{enumerate}
\item Defining covers $\MM$ such that there is no global section $s \in \EE(X)$ which 
satisfies the formula 
\[ \phi_{\MM} \; := \;\bigwedge_{C \in \MM} \XOR(C) . \]
\item Providing quantum representations for these covers, which interpret the measurements by quantum observables in such a way that \textit{every} quantum model for this set of observables has its support included  in $\XOR(C)$ for each $C \in \MM$, and hence is strongly contextual.
\end{enumerate}

We shall explain the quantum aspects in a later section. Here we shall  investigate the combinatorial structures involved in the first part.

We shall  give a simple combinatorial condition on the cover $\MM$ which is implied by the existence of a global section $s$ satisfying  $\phi_{\MM}$. Violation of this condition therefore suffices to prove that no such global section exists.

For each $m \in X$, we define
\[ \MM(m) := \{ C \in \MM \mid m \in C \} . \]
\begin{proposition}
If $\phi_{\MM}$ has a global section, then every common divisor of $\{ \card{\MM(m)} \mid m \in X \}$ must divide $\card{\MM}$.
\end{proposition}
\begin{proof}
Suppose there is a global section $s : X \rarr \{ 0, 1 \}$ satisfying $\phi_{\MM}$. Consider the set $X' \subseteq X$ of those $m$ such that $s(m) = 1$. Exactly one element of $X'$ must occur in every $C \in \MM$. Hence there is a partition of $\MM$ into the subsets $\MM(m)$ indexed by the elements of $X'$. Thus
\[ \card{\MM} = \sum_{m \in X'} \card{\MM(m)} . \]
It follows that, if there is a common divisor of the numbers $\card{\MM(m)}$,  it must divide $\card{\MM}$.
\end{proof}

For example, if every $m \in X$ appears in an even number of  elements of $\MM$, while $\MM$ has an odd number of elements, then $\phi_{\MM}$ has no global section.
This corresponds to the `parity proofs' which are often used in verifying Kochen-Specker-type results \cite{cabello1996bell,WaegellAravind2011}.

The simplest example of this situation is the `triangle', \ie the measurement cover with  elements
\[  \{a, b\}, \{ b, c \}, \{ a, c \}  . \]
This example has also been discussed, in a somewhat different context, in \cite{liang2010specker}.

An example where $X$ has 18 elements, and there are 9 maximal compatible sets,  each with four elements, such that each element of $X$ is in two of these, appears in the 18-vector proof of the Kochen-Specker Theorem in \cite{cabello1996bell}.

\subsection{Kochen-Specker Graphs}
\label{graphsubsec}

The measurement covers which can be represented by quantum systems are of a particular form: they are generated by a \emph{symmetric binary compatibility relation}, since compatibility in quantum mechanics means that the observables pairwise commute. Thus, for example, the  `triangle' cannot arise from quantum observables.

This suggests that we should take account of this feature.
It turns out that this leads us directly to some standard notions in graph theory.

An undirected graph $G$ is specified by a finite set of vertices $V_G$, and a set of edges $E_G$, which are two-element subsets of $V_G$. A \emph{clique} of $G$ is a set $C \subseteq V_G$ with an edge between every pair of vertices in $C$. The set of maximal cliques of $G$ forms a measurement cover $\MG$.

Let $G$ be a graph. A set $S \subseteq V_G$ is called a \emph{stable transversal} \cite{berge1984strongly} if for every maximal clique $C$ of $G$ (\ie for every $C \in \MG$), $\card{S \cap C} = 1$. Note that it is necessarily the case that a stable transversal is \emph{independent}, \ie there is no edge between any pair of elements of $S$, since otherwise we could extend this pair to a maximal clique containing both.

\begin{proposition}
Let $G$ be a graph.
The formula $\phi_{\MM}$ defined on $\MG$ has a global section if and only if $G$ has a stable transversal.
\end{proposition}
\begin{proof}
Suppose $\phi_{\MM}$ has a global section $s$. Then  $T := \{ m \in X \mid s(m) = 1 \}$ is a stable traversal of $G$.

Conversely, suppose that $T$ is a stable transversal of $G$. If we define $s$ as the characteristic function of $T$ on $X$, then $s \models \XOR(C)$ for each maximal clique $C$ of $G$, and so $\phi_{\MM}$  has a global section.
\end{proof}

In order to apply graph-theoretic results to the quantum situation, we need to know which graphs can arise from families of quantum observables.
For reasons which will be explained when we discuss quantum representations in Section~\ref{qrep},
we are interested in graphs which can be labelled by vectors in $\Real^d$, such that two vertices are adjacent if and only if the corresponding vectors are orthogonal. It turns out that in graph theory, the complementary notion is used  \cite{lovasz1989orthogonal}, so we shall say that such graphs have a \emph{faithful orthogonal co-representation} in $\Real^d$.
We must also require  that the maximal cliques all have size $d$. 

Thus we define \emph{Kochen-Specker graphs} to be finite graphs $G$ such that:
\begin{itemize}
\item $G$ has a faithful orthogonal co-representation in $\Real^d$.
\item The maximal cliques of $G$ all have the same size $d$.
\item $G$ has no stable transversal.
\end{itemize}

Any such graph generates a measurement cover $\MG$ such that the formula $\phi_{\MG}$ has no global section; and every such graph can be realized by  quantum observables, as will be shown in Section~\ref{qrep}. Thus, these graphs provide explicit finite witnesses for generic strong contextuality.
An example is provided by the  orthogonality graph for $\Real^4$ defined by the set of 18 vectors given in \cite{cabello1996bell}, as well as the various sets of 31 or more vectors which have been found in $\Real^3$ \cite{kochen1975problem,peres1991two,bub1996schŸtte,pavi?i?2005kochen}.

A final desideratum is to provide a purely graph-theoretic condition for the existence of a faithful orthogonal co-representation.
In  \cite{lovasz1989orthogonal,lov‡sz2000correction} the following result is proved.
\begin{theorem}
Every graph on $n$ nodes whose complementary graph is $(n - d)$-connected  has a faithful orthogonal co-representation in $\Real^d$.
\end{theorem}

We note that a general graph-theoretic approach to contextuality, on somewhat different lines to ours, has been developed in \cite{cabello2010non}. Interesting connections are shown  between contextuality and Lovasz's $\vartheta$-function \cite{lov‡sz1979shannon}.

In \cite{cabello2010non},  `compatibility structures' are studied as sets of \emph{events} rather than measurements. 
This leads into the  formalism of \emph{convex operational theories} as `generalized probability theories'. By contrast, we distinguish between measurements, outcomes and events. This allows the functorial presheaf structure of probabilistic models to be articulated. Moreover, we use standard probability theory, as encapsulated in the distributions functor $\DP$. The non-classical, contextual features of models arise from their functorial variation over contexts. At the same time, this mathematical structure directly reflects the basic operational scenario described at the beginning of Section~2.
%This offers a mathematically well-structured and conceptually compe

\section{Global Sections and Hidden Variables}
\label{gsechvsec}

We shall now consider a general notion of hidden-variable model, and show that an empirical model is realized by a factorizable hidden-variable model if and only if it has a global section.

We are given a measurement cover $\MM$. 
We fix a set $\Lambda$ of values for a hidden variable. A hidden-variable model $h$ over $\Lambda$ assigns, for each $\lambda \in \Lambda$ and $C \in \MM$, a distribution $\hcl \in \DR\EE(C)$.
It also assigns a distribution $\hl \in \DR(\Lambda)$ on the hidden variables.
Note that this distribution is independent of the context; this is the standard structural assumption of \emph{$\lambda$-independence} \cite{dickson1999quantum}.
We require that for each $\lambda \in \Lambda$, the family $\{ \hcl \}_{C \in \MM}$ is compatible, \ie
for all $C, C' \in \MM$:
\[ \hcl | C \cap C' = \hcpl | C \cap C' . \]
Just as compatibility for empirical models corresponds to no-signalling, compatibility for hidden-variable models corresponds to the \emph{parameter independence} condition \cite{jarrett1984physical,shimony1986events}. 

We say that a hidden-variable model $h$ \emph{realizes} an empirical model $e$ if the probabilities specified by $e$ are recovered by averaging over the values of the hidden variable.
Formally, this says that for all $C \in \MM$ and $s \in \EE(C)$:
\[ e_C(s) \;\; = \;\; \sum_{\lambda \in \Lambda} \hcl(s) \cdot \hl(\lambda) . \]

The intended purpose of hidden-variable models is to explain the non-intuitive behaviour of empirical models, in particular those arising from quantum mechanics, by showing that it can be reproduced by a model whose behaviour is more intuitive, at the cost of introducing hidden variables. In particular, one would like to explain the non-local and contextual behaviour predicted by quantum mechanics in this way.
The general property which a hidden-variable model should satisfy in order to provide such an explanation is \emph{factorizability}, which subsumes both Bell locality \cite{bell1964einstein}, and a form of non-contextuality at the level of distributions.
It is defined as follows.

We say that a hidden-variable model $h$ is \emph{factorizable} if, for every $C \in \MM$, and $s \in \EE(C)$:
\[ \hcl(s) \; = \; \prod_{m \in C} \hcl | \{ m \}  (s | \{ m \}) . \]
This says that the probability assigned to a joint outcome factors as the product of the probabilities assigned to the individual measurements.
Note in particular that, if $m \in C \cap C'$, then the compatibility condition on $h$ implies that
$\hcl | \{ m \}  = \hcpl | \{ m \}$. Thus the probability distributions on outcomes of individual measurements are independent of the contexts in which they occur.

For Bell-type scenarios, factorizability corresponds exactly  to Bell locality \cite{bell1964einstein}. More generally, it asserts non-contextuality at the level of distributions.

Our main result can now be stated as follows.

\begin{theorem}
\label{equivth}
Let $e$ be an empirical model defined on a measurement cover $\MM$ for a distribution functor $\DR$. The following are equivalent.
\begin{enumerate}
\item $e$ has a realization by a factorizable hidden-variable model.
\item $e$ has a global section.
\end{enumerate}
\end{theorem}
\begin{proof}
Proposition~\ref{globsecdhvprop} shows that \textit{(ii)} implies \textit{(i)}. It remains to prove the converse.

Suppose that $e$ is realized by a factorizable hidden-variable model $h$.
For each $m \in X$, we define $\hlm := \hcl | \{ m \} \in \DR\EE(\{ m \})$ for any $C \in \MM$ such that $m \in C$. By the compatibility of the family $\{ \hcl \}$, this definition is independent of the choice of $C$. Also, we shall write $s | m $ rather than $s | \{ m \}$. We  define a distribution $\hlx \in \DR\EE(X)$ for each $\lambda \in \Lambda$, by:
\[ \hlx(s) = \prod_{m \in X} \hlm (s | m ) . \]
We must show that this is a distribution. We enumerate the set of measurements $X$ as $X = \{ m_1 , \ldots , m_p \}$. A global assignment $s \in \EE(X)$ can be specified by a tuple $(o_1, \ldots , o_p)$, where $o_i = s(m_i)$. Now we can calculate:
\[ \begin{array}{lcl}
& & \sum_{s \in \EE(X)} \prod_{m \in X} \hlm(s | m) \\
& = & \sum_{o_1 , \ldots , o_p} \prod_{i=1}^p \hlim (s | m_i) \\
& = & \sum_{o_1} h^{\lambda}_{m_1}(m_1 \mapsto o_1) \cdot (\sum_{o_2} h^{\lambda}_{m_2}(m_2 \mapsto o_2) \cdot ( \cdots ( \sum_{o_p} h^{\lambda}_{m_p}(m_p \mapsto o_p)) \cdots )) \\
& = & \sum_{o_1} h^{\lambda}_{m_1}(m_1 \mapsto o_1) \cdot (\sum_{o_2} h^{\lambda}_{m_2}(m_2 \mapsto o_2) \cdot ( \cdots ( 1 ) \cdots )) \\
& = & \qquad \vdots \\
& = & \sum_{o_1} h^{\lambda}_{m_1}(m_1 \mapsto o_1) \cdot 1 = 1.
\end{array}
\]
We now show that for each context $C$ in $\MM$, $\hlx | C = \hcl$.  We choose an enumeration of  $X$ such that $C = \{ m_1 , \ldots , m_q \}$, $q \leq p$.
\[ \begin{array}{lcl}
\hlx | C(s) 
& = & \sum_{s' \in \EE(X), s'|C = s} \hlx(s') \\
& = & \sum_{(o_1, \ldots , o_p), s = (o_1, \ldots , o_q)} \prod_{i=1}^p h^{\lambda}_{m_i} (m_i \mapsto o_i) \\
& = & \prod_{i=1}^q h^{\lambda}_{m_i} (m_i \mapsto s(m_i)) \cdot (\sum_{o_{q+1}, \ldots , o_p} \prod_{j = q+1}^p h^{\lambda}_{m_j} (m_j \mapsto o_j)) \\
& = & \hcl(s) \cdot 1 \; = \; \hcl(s).
\end{array}
\]
Now we define a distribution $d \in \DR\EE(X)$ by averaging over the hidden variables:
\[ d(s) \; := \; \sum_{\lambda \in \Lambda} \hlx(s) \cdot \hl(\lambda) . \]
We verify that this is a distribution:
\[ \begin{array}{lcl}
\sum_{s \in \EE(X)} d(s) & = & \sum_{\lambda \in \Lambda} \sum_{s \in \EE(X)} \hlx(s) \cdot \hl(\lambda) \\
& = & \sum_{\lambda \in \Lambda} \hl(\lambda) \cdot (\sum_{s \in \EE(X)} \hlx(s)) \\
& = & \sum_{\lambda \in \Lambda} \hl(\lambda) \cdot 1 = 1.
\end{array}
\]
It remains to show that $d$ restricts at each context $C$ to yield $e_C$.
\[ \begin{array}{lcl}
d | C (s) & = & \sum_{s' \in \EE(X), s'|C = s} d(s') \\
& = & \sum_{s' \in \EE(X), s'|C = s} \sum_{\lambda \in \Lambda} \hlx(s') \cdot \hl(\lambda) \\
& = & \sum_{\lambda \in \Lambda}  \hl(\lambda) \cdot \hlx | C (s) \\
& = & \sum_{\lambda \in \Lambda}  \hl(\lambda) \cdot \hcl (s) \\
& = & e_C(s) .
\end{array}
\]
Thus $d$ is a global section for $e$.
\end{proof}

This result provides a definitive justification for equating the phenomena of non-locality and contextuality with obstructions to the existence of global sections.

\section{Quantum Representations}
\label{qrep}

Since our aim is to investigate \textit{general properties} of systems and physical theories, it has been important that our entire discussion has been conducted without presupposing quantum mechanics, Hilbert spaces, etc. The mathematical structures which we \textit{have} used have arisen in a rather transparent fashion from the basic experimental scenario with which we began.

However, it is important to make explicit how the structures we have described can be represented in quantum mechanics.

We begin with measurement covers.
A quantum representation of a measurement cover on a set $X$ can be described as follows.
We fix a Hilbert space $\HH$. As usual, an \emph{observable}  is a bounded self-adjoint operator $A$ on $\HH$. Two observables $A$, $B$ are \emph{compatible} if they commute: $AB = BA$. In this case, the composite $AB$ is again self-adjoint, and hence forms an observable.

Given a set $\XX = \{ A_x \}_{x \in X}$ of observables on $\HH$  indexed by $X$, we form a measurement cover by taking 
$\MM$ to be the set of all maximal commuting subsets of $\XX$.
Note that pairwise commutation implies that the observables in each such subset, composed in any order, form a well-defined observable.
We say that an abstract measurement cover $\MM$ has a \emph{quantum representation} if it arises in this way.

For Bell-type scenarios, a quantum representation will have a particular form, reflecting the usual idea that the measurements are performed at a number of different sites, which may be space-like separated. We will have a family $\{ \HH_i \}$ of Hilbert spaces, one for each part. The elements of $X_i$ will index a family $\XX_i$ of incompatible (\ie non-commuting) observables on $\HH_i$. We make these into \emph{local observables} on the \emph{compound system} $\HH = \HH_1 \otimes \cdots \otimes \HH_k$, by defining
$A^i := I \otimes \cdots \otimes A \otimes \cdots \otimes I$ for each $A \in \XX_i$. Then $A^i$ commutes with $B^j$ whenever $i \neq j$, and we can form a measurement cover of Bell type on the compound system.

It is interesting in this connection to discuss a result due to Tsirelson \cite{tsirelson2006bell}.
This result can be stated, following \cite{scholz2008tsirelson}, as follows:
\begin{theorem}
\label{Tsirelthm}
Let $\{ X_i \}$ and $\{ Y_j \}$ be finite, commuting sets of positive operators on a Hilbert space $\HH$, generating finite-dimensional von Neumann sub-algebras of $\BB(\HH)$. Then there exist finite-dimensional Hilbert spaces $\HH_1$ and $\HH_2$ such that $\{ X_i \}$ can be mapped isomorphically into the subalgebra of operators on $\HH_1 \otimes \HH_2$ of the form $A \otimes I$, and $\{ Y_j \}$ can be mapped isomorphically into the subalgebra of operators of the form $I \otimes B$.
\end{theorem}

The import of this result is that, under the stated assumptions, tensor product structure can be retrieved automatically as a special case of the general situation of commuting operators on a single space. Thus the special form of representation for Bell-type scenarios is not really necessary, although it is the one which is standardly used.

Now we turn to events.
For simplicity, we shall confine ourselves to the finite-dimensional case. Recall that a self-adjoint operator $A$ has a \emph{spectral decomposition} 
\[ A = \sum_{i \in I} \alpha_i \Proj_i \]
where $\alpha_i$ is the $i$'th eigenvalue, and $\Proj_i$ is the projector onto the corresponding eigenspace. Measuring a quantum state $\rho$ with this observable will result in one of the observable outcomes $\alpha_i$, with probability $\Tr(\rho\Proj_i)$, while the state will be projected into the corresponding eigenspace.

For simplicity of notation, we shall focus on \emph{dichotomic quantum observables}, \ie self-adjoint operators on a Hilbert space $\HH$ with a spectral resolution into two orthogonal subspaces. In this case, we can use a standard two-element set $O = \{ 0, 1 \}$ to label these outcomes, and  the sheaf $\EE$ to record the collective outcomes of a compatible set of observables.

Thus for each basic measurement label $m$ in $X$, we have an observable $A_m$ with spectral decomposition $A_m = \amp \Prmp + \amn \Prmn$, where $\Prmp + \Prmn = I$. Given a maximal set of commuting observables $C = \{ A_{m_1}, \ldots , A_{m_k} \}$, for each $s \in O^C$ we have a projector $\Ps = \Pms{1}  \cdots  \Pms{k}$. The composed observable $A_C = A_{m_1}  \cdots  A_{m_k}$ has a decomposition of the form
\[ A_C = \sum_{s \in O^C} \alpha_{s} \Ps , \]
where $\alpha_{s} = \prod_{i} \alpha_{m_i}^{s(m_i)}$. 
To ensure that these eigenvalues can be associated with distinct outcomes, we need that $\alpha_s = \alpha_t$ implies that $s = t$. This can be achieved by appropriate choices of the eigenvalues $\alpha_m^i$.

It may well be the case that this decomposition contains redundant terms, in the sense that $\Ps = \mathbf{0}$ for some values of $s$. The important point is that these projectors do yield a resolution of the identity:
\[ \sum_{s \in O^C} \Ps \; = \; (\Proj_{m_1}^{0} + \Proj_{m_1}^{1})  \cdots (\Proj_{m_k}^{0} + \Proj_{m_k}^{1}) \; = \; I^k \; = \; I . \]

Now we consider empirical models. Suppose we are given an empirical model $e$ on a measurement cover $\MM$, which has a quantum representation in the form described above, based on a Hilbert space $\HH$.
A quantum representation of  $e$ is given by a state $\rho$ on $\HH$. 
For each compatible set of observables $C \in \MM$, the state defines a probability distribution $\rho_C$ on $\EE(C)$, by the standard `statistical algorithm' of quantum mechanics:  $\rho_C(s) = \Tr(\rho \Ps)$.
Thus $\rho_C \in \mathcal{D}_{\Rpos}\EE(C)$ for each $C \in \MM$. 

An interesting point now arises: do the distributions $\{ \rho_C \}$ necessarily form a compatible family? In the case of a Bell-type scenario, the fact that they do is the content of the standard \emph{no-signalling theorem} \cite{ghirardi1980general}. However, Bell-type scenarios are very special cases of measurement covers. We shall therefore verify explicitly that the distributions determined by a quantum state, with respect to any family of sets of commuting observables, do  form a compatible family in the sense of sheaf theory. We can regard this as a generalized form of no-signalling theorem.

\begin{proposition}[Generalized No-Signalling]
\label{gnsigprop}
The family of distributions $\{ \rho_C \}$  on  families of commuting  observables  defined by a quantum state $\rho$ are compatible on overlaps: for all $C$, $C'$:
\[ \rho_C | C \cap C' \; = \; \rho_{C'} | C \cap C'  . \]
\end{proposition}
\vspace{0.01in}
\begin{proof}
Firstly, we define $C_0 := C \cap C'$, $C_1 := C \setminus C_0$, and $C_2 := C' \setminus C_0$. Thus $C$ is the disjoint union of $C_1$ and $C_0$, and $C'$ is the disjoint union of $C_2$ and $C_0$. Note that $\EE(C) \cong \EE(C_0) \times \EE(C_1)$, and $\EE(C') \cong \EE(C_0) \times \EE(C_2)$. Thus we can write $s \in \EE(C)$ as $s = (s_0, s_1)$, and similarly for sections in $\EE(C')$. In this notation, $\Proj_{(s_0, s_1)} = \Proj_{s_0}\Proj_{s_1}$.
Now we can calculate:
\[ \begin{array}{lcl}
\rho_C | C_0 (s_0) & = & \sum_{s_1 \in \EE(C_1)} \rho_{C}(s_0, s_1) \\
& = & \sum_{s_1 \in \EE(C_1)} \Tr(\rho \Proj_{(s_0, s_1)}) \\
& = & \sum_{s_1 \in \EE(C_1)} \Tr(\rho \Proj_{s_0} \Proj_{s_1}) \\
& = &  \Tr( \sum_{s_1 \in \EE(C_1)}  \rho \Proj_{s_0} \Proj_{s_1}) \\
& = &  \Tr( \rho \Proj_{s_0} \sum_{s_1 \in \EE(C_1)}  \Proj_{s_1}) \\
& = &  \Tr(\rho \Proj_{s_0} I) \\
& = &  \Tr(\rho \Proj_{s_0} ) \\
& = &  \rho_{C_0}(s_0) .
\end{array}
\]
A similar computation shows that $\rho_{C'} | C_0 (s_0) =  \rho_{C_0}(s_0)$. Hence 
$\rho_C | C \cap C' \; = \; \rho_{C'} | C \cap C'$.
\end{proof}

Thus we see that quantum mechanics obeys a general form of no-signalling, which applies to  compatible families of observables in general, not just those represented as operating on different factors of a tensor product.
%, and hence considered as possibly space-like separated.
This form of no-signalling says that, \textit{at the level of distributions}, the statistics obtained for a measurement on a given state are independent of the context of other compatible measurements which may also have been performed.

We can in fact use Tsirelson's Theorem~\ref{Tsirelthm} to reduce this result to the standard form of no-signalling theorem.\footnote{We thank one of the journal referees for this observation.}
Given sets of commuting observables $C$ and $C'$, every operator in $C \cap C'$ commutes with every operator in the symmetric difference $C \nabla C'$, so  Theorem~\ref{Tsirelthm} applies, and we can represent these two sets of observables as acting on different factors of a tensor product. The standard version of no-signalling can now be used to show that the marginals on the first factor are independent of the choice of measurement on the second.

\subsection{GHZ Models}
We shall briefly review how GHZ models, which were used in Section~\ref{scsec}, can be represented in quantum mechanics.
For $n > 2$, we take the Hilbert space to be the tensor product of $n$ qubit spaces.
The local observables in each factor are the $X$ and $Y$ spin measurements, represented in the $Z$ basis by eigenvectors for spin Right or Left along the $x$-axis, with basis vectors 
\[ \frac{| \ua  \rangle + | \da \rangle}{\sqrt{2}}, \qquad \frac{| \ua  \rangle - | \da \rangle}{\sqrt{2}} \]
and similarly for spin Forward or Back along the $y$-axis, with basis vectors
\[ \frac{| \ua  \rangle + i | \da \rangle}{\sqrt{2}}, \qquad \frac{| \ua  \rangle - i | \da \rangle}{\sqrt{2}} . \]
We shall label the outcomes as $0$ for spin Right for $X$ and spin Forward for $Y$; and $1$ for spin Left and spin Back respectively. 

The model is then generated by the GHZ state \cite{greenberger1989going,greenberger1990bell}, written in the $Z$ basis as
\[ \frac{\mid \uparrow \cdots \uparrow  \rangle \; + \; \mid \downarrow \cdots \downarrow \rangle}{\sqrt{2}} . \]
If we measure each particle with a choice of $X$ or $Y$ observable,  the probability for each outcome  is given by 
the square modulus of 
the inner product
\[ | \langle \mbox{GHZ}  \mid b_1 \cdots b_n \rangle |^2 ,  \]
where $b_i$ is the basis vector corresponding to the given outcome in the $i$'th component.

This computation is controlled by the product of the $\ket{\da}$-coefficients of the basis vectors, and hence by the 
cyclic group of order 4 generated by $i$. 

The following table gives the coefficients of the  $| \da \rangle$ components labelled by measurement and outcome:
\begin{center}
\begin{tabular}{l|cc}
& 0 & 1    \\ \hline
$X$ & $+1$ & $-1$  \\
$Y$ & $+i$ & $-i$  \\
\end{tabular}
\end{center}

The probability table for this model can be specified as follows:
\begin{itemize}
\item Each row with an odd number of $Y$ measurements has full support.
\item Each row in which the number of $Y$ measurements is a multiple of $4$ has as support those entries with an even number of outcomes labelled $1$.
\item Each row in which the number of $Y$ measurements is $4k + 2$ has as support those entries with an odd number of  outcomes labelled $1$.
\item In each case, the distribution is uniform on the support.
\end{itemize}
Note that by `row' here we mean a row of the probability table, \ie the distribution on the set of sections over a given measurement context.

\noindent Thus the interesting structure of this model arises purely from the support.

\subsection{Kochen-Specker Representations}
\label{KSqrep}

We shall now discuss how the abstract discussion of Kochen-Specker situations in Section~\ref{abskssec} can be represented in terms of quantum mechanics.

We shall consider a particular form of dichotomic observables. Given unit vectors $\ee_1, \ldots , \ee_k$ representing distinct rays in a Hilbert space $\HH$, we write
\[ \Aej := j \cdot \Prej + 0 \cdot \Pnj . \]
Then we can take $\XX = \{ \Aeone , \ldots , \Aek \}$
as a set of measurements. Note that $\Aei$ commutes with $\Aej$ if and only if $\ee_i$ is orthogonal to $\ee_j$. Also, the composition of a set of commuting observables $\{ \Aei \}_{i \in I}$ will have a spectral decomposition of the form
%\begin{equation}
\[
\label{dichobseq}
\sum_{i \in I}  i \cdot \Prei \; + \; 0 \cdot \Proj_{\{\ee_i \mid i \in I \}^{\bot}} . 
\]
%\end{equation}
If we measure any state with this observable, the outcome must be that we get exactly one of the branches $\Prei$, with eigenvalue $i$; or that we get `none of the above', corresponding to the branch $\Proj_{\{\ee_i \mid i \in I \}^{\bot}}$, with eigenvalue $0$. Moreover, if the cardinality of $I$ equals the dimension of the Hilbert space, then the latter case cannot apply.

If we now consider how outcomes are represented in the sheaf $\EE$, we see that we indeed have an \textit{a priori}~condition on those sections $s$ which can be in the support of a distribution coming from a quantum state, as desired. Namely, using $x_i$ as a label for $\Aei$, and taking $s(x_i) = 1$ for the outcome corresponding to $\Prei$ for this observable, we see that the only sections which are possible are those which assign $1$ to at most one measurement. Moreover, for those sets of compatible observables whose cardinality equals the dimension of the space --- which must necessarily be maximal, and hence will appear in the measurement cover --- exactly one measurement must be assigned $1$.

Thus if we take a set of unit vectors indexed by $X$, such that each vector is contained in at least one orthonormal basis indexed by a subset of $X$, the measurement cover $\MM$ represented by the observables $A_x$ will have the following key property:
for \textit{any} quantum state $\rho$, the support of the corresponding empirical model will satisfy the formula $\XOR(C)$ for each context $C$ in $\MM$. So the problem of exhibiting a state-independent form of strong contextuality has been reduced to the problem of finding a Kochen-Specker graph, as described in Section~\ref{abskssec}.

\subsection{Bell-Type Scenarios and Kochen-Specker Theorems}

The measurement covers arising from Bell-type scenarios are a rather special class, which can be characterized as follows.

\begin{proposition}
A measurement cover $\MM$ arises from a Bell-type scenario if and only if it is the family of maximal cliques of a graph $G = (X, E_G)$ which is the complement of an equivalence relation $R$ on $X$:
\[ E_G = \{ \{ x, y \} \mid \neg (xRy) \} . \]
\end{proposition}
\vspace{0.01in}
\begin{proof}
Equivalence relations are in bijective correspondence with partitions $X = \coprod_i X_i$.
The maximal cliques of $G$ are exactly the \emph{transversals} of this partition, \ie the sets $T \subseteq X$ such that $T$ intersects with each $X_i$ in exactly one element.
\end{proof}

Note in particular that in Bell-type scenarios, \textit{incompatibility is transitive}. This is by no means the general case. In terms of operators, $A$ and $B$ may commute, while $C$ may fail to commute with either.

The more complex configurations typical of Kochen-Specker constructions can never arise from Bell-type scenarios.

\begin{proposition}
Consider a measurement cover $\MM$ of Bell type, and any quantum representation of $\MM$.
For any $s \in \EE(C)$ with $C \in \MM$, there is a quantum state $\rho$ such that $s$ is in the support of $\rho_C$. 
\end{proposition}
\begin{proof}
Given $s$, we define the local state $\rho_i := \ket{\psi_i}\bra{\psi_i}$ for each $i$, where $\psi_i$ is the eigenvector corresponding to the outcome specified by $s$ for the measurement at $i$.
Then the model defined by the state $\rho := \rho_1 \otimes \cdots \otimes \rho_n$ has $s$ in its support.
\end{proof}

Hence there is no Kochen-Specker-type theorem for Bell-type scenarios. While, as we have seen, there \textit{are} model-specific strong contextuality results, there are no \textit{generic} results. The measurement covers arising from these scenarios are simply not rich enough in their combinatorial structure of overlapping intersections to support  a result of this form.

\section{Postlude}
\label{postludesec}

Our treatment of non-locality and contextuality makes a number of points:

\begin{itemize}
\item Firstly, it is carried out at a high level of generality, and without any presupposition of quantum mechanics. None of the characteristic mathematical structures of quantum mechanics, such as complex numbers, Hilbert spaces, operator algebras, or projection lattices, are needed to expose the key structural issues. This characteristic is shared to some extent by other foundational approaches, such as generalized probabilistic theories \cite{barrett2007information}, but these formalisms are still rather closer to that of quantum mechanics, and indeed have been directly suggested by it. 
Structures such as sheaves and presheaves varying over contexts can be seen as basic elements of a general `logic of contextuality', and related structures have been used for a wide range of purposes, e.g. in the semantics of computation \cite{olestype,tennent1986functor,cattani1997presheaf}. This opens up the possibility of making some interesting connections between the study of non-locality and contextuality in physics, and ideas arising in other fields.

%\item Non-locality and contextuality are studied in a unified framework throughout.
%While other work such as \cite{cabello2010non} also relates the two, our approach is particularly systematic; 
%our definitions and results specialize to yield standard formulations of either as special cases, but subsume both.
%Unless causality is represented explicitly, non-locality should be seen as a special case of contextuality.

\item The sheaf-theoretic language, which directly captures the idea of structures varying over contexts, is a canonical setting for studying contextuality. Moreover, as we have seen, the gluing conditions and the existence of global sections captures the essential content of non-locality and contextuality in a canonical mathematical form.

This opens the door to the use of the powerful methods of sheaf theory, which plays a major r\^ole in modern mathematics, in analyzing the structure of non-locality and contextuality.
These notions are still poorly understood in multipartite and higher-dimensional settings.

In \cite{abrmansfield11}, the first author, with Shane Mansfield  and Rui Soares Barbosa, define an abelian presheaf based on the support of an empirical model. The \u{C}ech cohomology of this presheaf with respect to the measurement cover is used to define a cohomological obstruction to locality or non-contextuality, as a certain cohomology class. It is shown for a number of salient examples, including PR boxes, GHZ states, the Peres-Mermin square, and the 18-vector configuration from \cite{cabello1996bell} giving a proof of the Kochen-Specker theorem in four dimensions, that this obstruction does not vanish, thus yielding cohomological witnesses for contextuality.
While these results are preliminary, they suggest that the use of cohomological methods in the study of non-locality and contextuality has some promise.

%We plan to explore the applications of cohomological methods suggested by  this observation in future work.

\item The canonical form of description of the key concepts in terms of the existence of global sections largely replaces any explicit mention of hidden variables. These appear only in Section~\ref{gsechvsec}, in the context of a foundational result showing the equivalence of local hidden-variable realizations to the existence of global sections. On the other hand, empirical models, which can be seen as directly related to observation and experiment, play a prominent r\^ole throughout the paper.

\end{itemize}

There is also an interesting conceptual point to be made in relation to \emph{incompatibility of measurements}. Usually, this is taken to be a postulate of quantum mechanics, and specific to the quantum-mechanical formalism of non-commuting observables. However, in the light of general results such as those obtained in this paper, in a line of work going back to that of Fine \cite{fine1982hidden}, a different view emerges. The incompatibility of certain measurements can be interpreted as the impossibility --- in the sense of mathematically provable non-existence --- of joint distributions on \textit{all} measurements which marginalize to yield the observed empirical distributions. Thus, if we refer to the experimental scenario with which we began Section~2,
this shows that there cannot be, even in principle, any  such scenario in which all measurements can be performed jointly, which is consistent with the actually observed outcomes.

Thus the incompatibility of certain measurements is revealed as a theory-independent structural impossibility result for certain families of empirical distributions. These families include those predicted by quantum mechanics, and confirmed by experiment; but the result itself is completely independent of quantum mechanics. Thus in this sense, we can say that incompatibility is \textit{derived} rather than \textit{assumed}.

\subsection{Related Work}

The present paper builds on our previous work, in particular \cite{abramsky2010relational} by the first author, and \cite{BK,brandenburger2008classification} by the second author (with H. Jerome Keisler and Noson Yanofsky, respectively).
A natural direction for generalization of the results in the present paper would be from the finite setting considered here to the measure-theoretic one studied in  \cite{BK}; note that the distribution functor can be defined over general measure spaces \cite{giry1982categorical}.

Since we use sheaf theory as our mathematical setting, there is an obvious point of comparison with the topos approach, as developed by Isham, Butterfield, D\"oring, Heunen, Landsman, Spitters et al. \cite{isham1998topos,dšring2008topos,heunen2009topos}.

The general point that presheaves varying over a poset of contexts provides a natural mathematical setting for studying contextuality phenomena is certainly a common feature. It should also be mentioned that presheaves have been used for similar purposes in the context of the semantics of computation, e.g. in the Reynolds-Oles functor-category semantics for programs with state \cite{olestype,tennent1986functor}, and in the presheaf models for concurrency of Cattani and Winskel \cite{cattani1997presheaf}.

A more specific source of inspiration is the  important insight in \cite{isham1998topos}, which initiated the whole topos approach, that the Kochen-Specker theorem could be reformulated very elegantly in presheaf terms, as stating the non-existence of global sections of a certain presheaf. 

On the other hand, there are several differences between the present work and the topos approach. For example, the topos approach focusses on a specific structure, the spectral presheaf, based on an operator algebra.
In this sense, it uses concepts specific to quantum mechanics from the very start.
Moreover, many of the key structures introduced in our work, such as the distribution functor and measurement covers,  do not appear in the topos approach.
One of our central objectives is to give a unified account of contextuality and non-locality, but locality issues have not been considered in the topos approach; nor has extendability, another key topic for us. 
It will, of course, be interesting to see if additional commonalities develop in future work.

The appearance of negative probabilities in quantum mechanics has a long history.  The Wigner  quasi-probability distribution \cite{wigner1932quantum}, further developed by Moyal \cite{moyal1949quantum}, is a phase-space representation of quantum mechanics using negative probabilities.  Feynman views such negative probabilities as a calculational convenience \cite{feynman1987negative}.  He explains that the appearance of a negative probability for a certain outcome does not invalidate the theory being used.  Rather, this tells us that the relevant conditions cannot be realized, or that the outcome cannot be verified, or both.  More specifically related to what we do, Sudarshan and Rothman \cite{sudarshan1993new} show that a local hidden-variable analysis of the Bell model is possible, if certain values of the hidden variable arise with negative probability. Finally, in Dirac \cite{Dirac42}, negative probabilities enter in the relativistic extension of quantum mechanics.

\ack

We thank Noson Yanofsky for many important discussions and for encouragement throughout the development of the ideas in this paper.  Harvey Brown, Andreas D\"oring, Peter Johnstone, Aleks Kissinger, Ray Lal, Pierfrancesco La Mura, Elliot Lipnowski, Shane Mansfield, John Myers, Andrei Savochkin, Natalya Vinokurova, and audiences at several meetings, including the 2010--2011 Clifford Lectures, Tulane University, March 2011,  the LINT Workshop in Oxford, April 2011, the McGill University Bellairs Workshop, April 2011, The Knots in Washington Workshop, George Washington University, May 2011,  the Workshop on Topological Quantum Information, Pisa,  May 2011, and the Conference on Unconventional Computation, Turku, June 2011, provided valuable input.  Financial support from the Stern School of Business, the CUNY Graduate Center, EPSRC Senior Research Fellowship EP/E052819/1, and the U.S. Office of Naval Research Grant Number N000141010357 is gratefully acknowledged.

We thank the two anonymous journal referees for their comments, which have led to a number of clarifications and improved references to the literature. One of the referees in particular made a number of detailed and perceptive comments, which led to several significant improvements in the revised version of the paper.

\section*{Appendix}

Firstly, we recall some set-theoretic notations.

We write 
$\card{S}$ for the cardinality of a  set $S$. If $f : X \rarr Y$ is a function and $X' \subseteq X$, we write $f \fres X' : X' \rarr Y$ for the restriction of $f$ to $X'$. We write $Y^{X}$ for the set of functions from $X$ to $Y$, and $\Pow(X)$ for the powerset of $X$. 

A family of sets $\{ X_i \}_{i \in I}$ is \emph{disjoint} if $X_{i} \cap X_{j} = \vn$ whenever $i \neq j$. We write $\coprod_{i \in I} X_i$ for the union of a disjoint family.
Given a disjoint family $\{ X_i \}_{i \in I}$, there is an isomorphism
\[ \Pow(\coprod_{i \in I} X_i) \rTo^{\cong} \prod_{i \in I} \Pow(X_i) :: S \mapsto (S \cap X_i)_{i \in I} . \]

A \emph{category} has a collection of objects $A, B, C, \ldots$, and arrows $f, g, h, \ldots$.
Each arrow has  specified \emph{domain} and \emph{codomain} objects: notation is $f : A \rarr B$ for an arrow $f$ with domain $A$ and codomain $B$. Given arrows $f : A \rarr B$ and $g : B \rarr C$, we can form the \emph{composition} $g \circ f : A \rarr C$. Composition is associative, and there are identity arrows $\id_{A} : A \rarr A$ for each object $A$, with $f \circ \id_{A} = f$, $\id_{A} \circ g = g$, for every $f : A \rarr B$ and $g : C \rarr A$.

Our main examples of categories will be $\Set$, with sets as objects and functions as arrows; and partially ordered sets $(P, {\leq})$, where there is a single arrow from $p$ to $q$ if $p \leq q$, and none otherwise. The \emph{opposite category} $P^{\op}$ is the category formed from the opposite poset $(P, {\geq})$.

If $\CC$ and $\DD$ are categories, a \emph{functor} $F : \CC \lrarr \DD$ assigns an object $FA$ of $\DD$ to each object $A$ of $\CC$; and an arrow $Ff : FA \rarr FB$ of $\DD$ to every arrow $f : A \rarr B$ of $\CC$.
These assignments must preserve composition and identities: $F(g \circ f) = F(g) \circ F(f)$, and $F(\id_{A}) = \id_{FA}$.

A \emph{presheaf} on a poset $P$ is a functor $P^{\op} \lrarr \Set$.

\section*{References}

\bibliographystyle{unsrt}

\bibliography{bdbib}

\end{document}

%% file: scenario.tex
\begin{tikzpicture}[scale=2.54]
% dpic version 2011.01.25 option -g for TikZ and PGF 1.01
\ifx\dpiclw\undefined\newdimen\dpiclw\fi
\global\def\dpicdraw{\draw[line width=\dpiclw]}
\global\def\dpicstop{;}
\dpiclw=0.8bp
\dpiclw=2bp
\dpicdraw[fill=lightgray](0,-0.321429) rectangle (0.964286,0.321429)\dpicstop
\dpiclw=1bp
\dpicdraw (0.482143,-0.559286) circle (0.050619in)\dpicstop
\draw (0.482143,-0.430714) node[above=-1.928571bp]{$a$};
\draw (0.610714,-0.559286) node[right=-1.928571bp]{$b$};
\draw (0.482143,-0.687857) node[below=-1.928571bp]{$c$};
\draw (0.353571,-0.559286) node[left=-1.928571bp]{$d$};
\draw (0.482143,-0.559286) node{$\mathbf{\cdot}$};
\dpicdraw[fill=black](0.424286,-0.617143) circle (0.010124in)\dpicstop
\dpicdraw[fill=lightgray](2.25,-0.321429) rectangle (3.214286,0.321429)\dpicstop
\dpicdraw (2.732143,-0.559286) circle (0.050619in)\dpicstop
\draw (2.732143,-0.430714) node[above=-1.928571bp]{$a$};
\draw (2.860714,-0.559286) node[right=-1.928571bp]{$b$};
\draw (2.732143,-0.687857) node[below=-1.928571bp]{$c$};
\draw (2.603571,-0.559286) node[left=-1.928571bp]{$d$};
\draw (2.732143,-0.559286) node{$\mathbf{\cdot}$};
\dpicdraw[fill=black](2.79,-0.501429) circle (0.010124in)\dpicstop
\dpicdraw (0.160714,0.353571)
 ..controls (0.160714,0.531092) and (0.304623,0.675)
 ..(0.482143,0.675)
 ..controls (0.659663,0.675) and (0.803571,0.531092)
 ..(0.803571,0.353571)\dpicstop
\dpicdraw (0.739286,0.353571)
 --(0.803571,0.353571)\dpicstop
\dpicdraw (0.733667,0.407034)
 --(0.796547,0.4204)\dpicstop
\dpicdraw (0.717055,0.458161)
 --(0.775782,0.484308)\dpicstop
\dpicdraw (0.690176,0.504716)
 --(0.742184,0.542502)\dpicstop
\dpicdraw (0.654205,0.544666)
 --(0.697221,0.592439)\dpicstop
\dpicdraw (0.610714,0.576264)
 --(0.642857,0.631937)\dpicstop
\dpicdraw (0.561604,0.598129)
 --(0.58147,0.659268)\dpicstop
\dpicdraw (0.509022,0.609306)
 --(0.515741,0.673239)\dpicstop
\dpicdraw (0.455264,0.609306)
 --(0.448544,0.673239)\dpicstop
\dpicdraw (0.402681,0.598129)
 --(0.382816,0.659268)\dpicstop
\dpicdraw (0.353571,0.576264)
 --(0.321429,0.631937)\dpicstop
\dpicdraw (0.310081,0.544666)
 --(0.267065,0.592439)\dpicstop
\dpicdraw (0.27411,0.504716)
 --(0.222102,0.542502)\dpicstop
\dpicdraw (0.247231,0.458161)
 --(0.188503,0.484308)\dpicstop
\dpicdraw (0.230619,0.407034)
 --(0.167738,0.4204)\dpicstop
\dpicdraw (0.225,0.353571)
 --(0.160714,0.353571)\dpicstop
\dpicdraw (0.482143,0.353571)
 --(0.638168,0.46693)\dpicstop
\filldraw (0.657061,0.440926)
 --(0.742184,0.542502)
 --(0.619274,0.492934) --cycle
\dpicstop
\dpicdraw (2.410714,0.353571)
 ..controls (2.410714,0.531092) and (2.554623,0.675)
 ..(2.732143,0.675)
 ..controls (2.909663,0.675) and (3.053571,0.531092)
 ..(3.053571,0.353571)\dpicstop
\dpicdraw (2.989286,0.353571)
 --(3.053571,0.353571)\dpicstop
\dpicdraw (2.983667,0.407034)
 --(3.046547,0.4204)\dpicstop
\dpicdraw (2.967055,0.458161)
 --(3.025782,0.484308)\dpicstop
\dpicdraw (2.940176,0.504716)
 --(2.992184,0.542502)\dpicstop
\dpicdraw (2.904205,0.544666)
 --(2.947221,0.592439)\dpicstop
\dpicdraw (2.860714,0.576264)
 --(2.892857,0.631937)\dpicstop
\dpicdraw (2.811604,0.598129)
 --(2.83147,0.659268)\dpicstop
\dpicdraw (2.759022,0.609306)
 --(2.765741,0.673239)\dpicstop
\dpicdraw (2.705264,0.609306)
 --(2.698544,0.673239)\dpicstop
\dpicdraw (2.652681,0.598129)
 --(2.632816,0.659268)\dpicstop
\dpicdraw (2.603571,0.576264)
 --(2.571429,0.631937)\dpicstop
\dpicdraw (2.560081,0.544666)
 --(2.517065,0.592439)\dpicstop
\dpicdraw (2.52411,0.504716)
 --(2.472102,0.542502)\dpicstop
\dpicdraw (2.497231,0.458161)
 --(2.438503,0.484308)\dpicstop
\dpicdraw (2.480619,0.407034)
 --(2.417738,0.4204)\dpicstop
\dpicdraw (2.475,0.353571)
 --(2.410714,0.353571)\dpicstop
\dpicdraw (2.732143,0.353571)
 --(2.603096,0.496892)\dpicstop
\filldraw (2.626983,0.5184)
 --(2.517065,0.592439)
 --(2.579209,0.475384) --cycle
\dpicstop
\draw (0.482143,-0.880714) node[below=-1.928571bp]{Alice};
\draw (2.732143,-0.880714) node[below=-1.928571bp]{Bob};
\draw (4.5,0) node{\ldots};
\end{tikzpicture}